\documentclass[spreprintnumbers,amsmath,amssymb]{revtex4}
\usepackage{graphicx}

\newcommand{\ee}{\end{eqnarray}}
\newcommand{\be}[1]{\begin{eqnarray} \mbox{$\label{#1}$} }
\newcommand{\mtrx}[2]{\left(\begin{array}{#1} #2 \end{array}\right)}
\newcommand{\eref}[1]{(\ref{#1})}
\newcommand{\der}{\mathrm{d}}
\newcommand{\tr}{\mathrm{Tr}\,}

\newcommand\ie {{\it i.e.~}}
\newcommand\eg {{\it e.g.~}}

\newcommand\etal{{\it et al.~}}

\newcommand{\ul}\underline

\newcommand{\Go}{{G^0}}
\newcommand{\GR}{G^\mathrm{R}}
\newcommand{\SR}{\Sigma^\mathrm{R}}
\newcommand{\GoR}{G^{0\mathrm{R}} }
\newcommand{\GA}{G^\mathrm{A}}
\newcommand{\SA}{\Sigma^\mathrm{A}}
\newcommand{\GoA}{G^{0\mathrm{A}} }

\newcommand{\GK}{G^\mathrm{K}}

\newcommand{\GL}{G^<}
\newcommand{\SL}{\Sigma^<}

\newcommand{\prip}[1]{\mathcal{P}\left(#1\right)}
\newcommand{\pripp}{\mathcal{P}_+}
\newcommand{\pripm}{\mathcal{P}_-}
\newcommand{\GLr}{{\GL}^\mathrm{r}}
\newcommand{\GLa}{{\GL}^\mathrm{a}}

\newcommand{\evu}{\mathcal{U}}

\newcommand{\DT}{\Delta}
\newcommand{\pr}{^\prime}

\renewcommand{\vec}[1]{\text{\boldmath{$ #1 $}}}
\newcommand{\uv}[1]{\vec{\hat{#1}}} 
\newcommand{\ksum}[1]{\int_{#1}}  
\newcommand{\ki}{\ksum{\vec{k\pr}}}

\newcommand{\Wkk}{W_{\vec{k}\vec{k\pr}}}

\newcommand{\fermi}{\mathrm{F}}
\newcommand{\rel}{\textrm{rel}}

\newcommand{\sch}{Schr\"odinger{ }}

\newcommand{\vf}{v_\mathrm{F}}

\newcommand{\feq}{f^\textrm{eq}}
\newcommand{\fee}{f^{(E)}}

\newcommand{\varomega}{{\tilde{\omega}}}
\newcommand{\erg}{\epsilon}
\newcommand{\aen}{{\erg_0}}
\newcommand{\ben}{b}
\newcommand{\cen}{c}
\newcommand{\sproj}{S}
\newcommand{\idm}{\mathbf{1}}
\newcommand{\ttr}{\tau_\textrm{tr}}

\newcommand{\vecvo}{\vec{v}_0}
\newcommand{\vkm}{\mathbf{v}}

\newcommand{\ordo}[1]{\mathcal{O}(#1)}
\newcommand{\FD}[1]{f_\mathrm{FD}\left(#1\right)}
\newcommand{\FDf}{f_\mathrm{FD}}

\newcommand{\DS}{D} 

\newcommand{\grex}{\mathcal{D}}

\newcommand{\Drive}{\mathcal{D}}
\newcommand{\Precession}{\mathcal{S}}

\newcommand{\JC}{\mathcal{J}}
\newcommand{\JCd}{\mathcal{J}^\delta}
\newcommand{\JCp}{\mathcal{J}^\mathrm{P}}
\newcommand{\JCpx}{\mathcal{J}^{\mathrm{P}X}}
\newcommand{\JCpy}{\mathcal{J}^{\mathrm{P}Y}}
\newcommand{\IC}{\mathcal{I}}

\renewcommand{\Re}{\mathrm{Re}\,}
\renewcommand{\Im}{\mathrm{Im}\,}

\newcommand{\Bkk}{\mathrm{B}}

\newcommand{\ME}{vN}
\newcommand{\GO}{G1}
\newcommand{\GT}{G2}
\renewcommand{\AA}{AA}
\newcommand{\SKBA}{SKBA}

\newcommand{\Htot}{H_\textrm{tot}}
\newcommand{\Ho}{H_0}
\newcommand{\Hi}{V}%{H_\textrm{i}}
\newcommand{\rhoI}{\rho^{\textrm{I}}} 
\newcommand{\INT}{^\textrm{I}} 

\newcommand{\nimp}{n_\textrm{imp}}
\newcommand{\singleimp}{u} 

\newcommand{\rhohat}{\hat{\rho}}
\newcommand{\rhohatrel}{\hat{\rho}_\textrm{rel}}

\newcommand{\qs}{q_\mathrm{s}}

\begin{document}

%\title{The importance of principal value terms and the sensitivity to choice of formalism in the Boltzmann treatment of conductivity in spin-orbit coupled systems}

%\title{Quantum corrections in the Boltzmann conductivity of graphene: \\
%importance of principal value terms  and sensitivity to the choice of formalism }

\title{Quantum corrections in the Boltzmann conductivity of graphene\\
and their sensitivity to the choice of formalism}
%:\\ importance of principal value terms and sensitivity to the choice of formalism}

\author{Janik Kailasvuori$^1$ and  Matthias C. L\"uffe$^2$}
\email {kailas@pks.mpg.de, lueffe@physik.fu-berlin.de,}

\affiliation{
$^1$Max-Planck-Institut f\"ur Physik komplexer Systeme, N\"othnitzer Stra\ss{}e 38, 01189 Dresden, Germany \\
$^2$Dahlem Center for Complex Quantum Systems \& Fachbereich Physik, Freie Universit\"at Berlin, Arnimallee 14, 14195 Berlin, Germany}

\date{\today}

\begin{abstract}
\noindent
Semiclassical spin-coherent kinetic equations can be derived from quantum theory with many different approaches (Liouville equation based approaches,
nonequilibrium Green's functions techniques, etc.). The  collision integrals turn out to be formally different, but coincide in textbook examples as well as for systems  where the spin-orbit coupling is only a small part of the kinetic energy like in related studies on the spin Hall effect.  In Dirac cone physics  (graphene, surface states of  topological insulators like $\textnormal{Bi}_{1-x}\textnormal{Sb}_x,\, \textnormal{Bi}_{2}\textnormal{Te}_3$ etc.), where this coupling constitutes the entire kinetic energy, the difference  
manifests itself in the  precise value of the electron-hole coherence originated quantum correction to the Drude conductivity $\sigma_0 \sim \tfrac{e^2}{h} \ell k_\fermi$. The leading correction is derived analytically for single and multilayer graphene with general scalar impurities. 
The often neglected principal value terms in the collision integral are 
important. Neglecting them 
yields a leading correction of order $(\ell k_\fermi)^{-1}$, whereas including them can give a correction of order $(\ell k_\fermi)^0$. 
 The latter  opens up a counterintuitive scenario with finite electron-hole coherent effects  at Fermi energies arbitrarily far above the neutrality point regime, for example in the form of a shift $\delta\sigma\sim \tfrac{e^2}{h}$ that only depends on the dielectric constant. 
 This residual conductivity, possibly related to the one observed in recent experiments,  depends crucially on the approach and could  offer a setting    
for experimentally singling out one of the  candidates.  
 Concerning the different formalisms we notice that the discrepancy between a density matrix approach and a Green's function approach is removed if  the Generalized Kadanoff-Baym Ansatz in the latter is replaced by an anti-ordered version. This issue of Ansatz may also be important for Boltzmann type treatments of  graphene 
  beyond linear response. 

\hspace{1cm}

 \end{abstract}

%\pacs{73.43.Cd, 71.10.Pm, 75.10.Pq}

\maketitle

\section{Introduction}
\noindent
Since the first isolation of graphene in 2004 \cite{Novoselov-Geim-etal-2004} the electrical conductivity of this system has attracted huge attention.  To a good approximation the electrons can be described as massless 2d  Dirac electrons for which the  spin-orbit interaction that  yields  
the characteristic Dirac cone  is given by the pseudospin derived from the bipartite honeycomb lattice.\cite{Wallace-1947,Semenoff-1984} The conical electron and hole bands touch at the  Dirac points. The Brillouin zone contains two inequivalent degenerate Dirac points---K and K'---that give an additional  valley index. Finally there is the ordinary electron spin.  To first approximation, the conductive properties of graphene involve only the pseudospin in a nontrivial way.

The main focus has been on undoped graphene, 
with the chemical potential exactly at the degenerate Dirac points. 
This regime of chemical potential close to zero---the Dirac regime---hosts the most exotic features, for example the finite
 conductivity minimum at seemingly zero charge carrier density.\cite{Novoselov-Geim-etal-2005a, Zhang-Kim-etal-2005a}  Quantum effects due to  electron-hole coherence (that is, pseudospin coherence)  like  Zitterbewegung can dictate the observed conductivity even to lowest approximation.\cite{Katsnelson-2006} We refer to  ref.~\cite{Castro-Neto-etal-review-2009} for a review on early work on the Dirac regime.  

Away from the Dirac regime, with a large enough charge carrier density
there is a crossover into the Boltzmann regime $\ell k_\fermi \gg 1$ (with $\ell$ the mean free path and $\hbar k_\fermi$ the Fermi momentum). Here, the conductivity can 
be understood to lowest order without 
taking into account quantum effects  such as  
 electron-hole coherence and are therefore more intuitive. The crossover between the two regimes has recently been studied numerically.\cite{Adam-Brouwer-DasSarma-2009, Cappelluti-Benfatto-2009,Trushin-etal-2010} 

\begin{figure}[htbp]
\begin{center}
\includegraphics[scale=0.8]{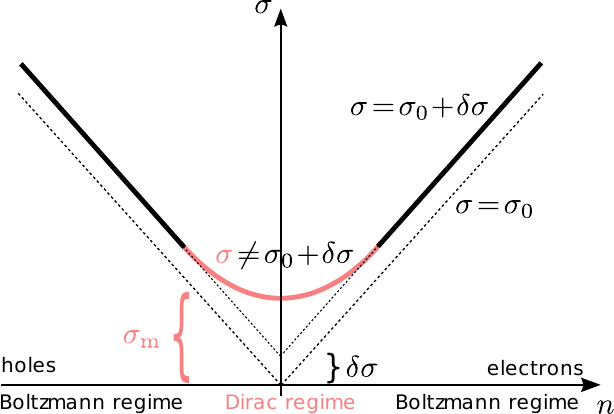}
\caption{ 
 Sketch of
the dc conductivity $\sigma$ in monolayer graphene as a function of the   electron density $n$ as observed in some experiments \cite{Novoselov-Geim-etal-2005a,
Zhang-Kim-etal-2005a,
Schedin-Geim-Morozov-Hill-Blake-Katsnelson-Novoselov-2007,
Tan-Zhang-Bolotin-Zhao-Adam-Hwang-DasSarma-Stormer-Kim-2007,Chen-Jang-Adam-Fuhrer-Williams-Ishigami-2008},
 in particular including a residual conductivity  $\delta \sigma$ as observed by Chen \etal\cite{Chen-Jang-Adam-Fuhrer-Williams-Ishigami-2008}. 
At the neutrality point $n=0$  the charge density is zero and  one would expect zero charge carrier density at low temperatures. One of the surprises of graphene is the  conductivity minimum
$\sigma_\textrm{m}\sim {e^2}/{h}$. In the Dirac regime $n\approx 0$ (red/gray) the usual criterion $\ell k_\fermi \gg1$ for a Boltzmann treatment is not satisfied {\bf (}at least not for screened charged impurities{\bf )}. Away from the Dirac regime 
a semiclassical approach should usually apply. The observed linear dependence in the Boltzmann regime (black) is described by the Drude conductivity $\sigma_0=2\,{e^2}\,\ell k_\fermi/h\propto |n|$ for 
screened charged impurities. (Point-like impurities, in contrast, yield $\sigma_0\propto |n|^0$. The resistivity due to them starts to compete with that of  screened charged impurities for $|n|$  large 
enough.) Effects of electron-hole coherence do not enter into the Drude conductivity but reveal themselves only in quantum corrections of higher order in $(\ell k_\fermi)^{-1}$. A contribution $(\ell k_\fermi)^{-1}$ can explain the initial convexity as one approaches the Dirac regime. 
A contribution $ (\ell k_\fermi)^0$ enters as  a constant shift in  the Boltzmann conductivity and contributes thus to  the residual conductivity $\delta \sigma$. The latter is  directly  read off by linear extrapolation.  A precise knowledge of other contributions to $\delta \sigma$ (\eg weak localization, \ldots),  allowing  precise estimate of  the  electron-hole coherent  contribution, which in turn would single out one of the many possible derivations discussed in the present paper. With one exception\cite{Trushin-etal-2010}  the contribution to the conductivity minimum depends also  in general on the approach. However, these differences can probably not be studied as cleanly as those in the residual conductivity.  
We want to stress that the residual conductivity is inherent  in  
the Boltzmann regime whereas the minimum conductivity is inherent  in  
the Dirac regime. }
\label{f:sketch}
\end{center}
\end{figure}

In many experiments on graphene samples on substrates  \cite{Novoselov-Geim-etal-2005a,
Zhang-Kim-etal-2005a,
Schedin-Geim-Morozov-Hill-Blake-Katsnelson-Novoselov-2007,
Tan-Zhang-Bolotin-Zhao-Adam-Hwang-DasSarma-Stormer-Kim-2007,Chen-Jang-Adam-Fuhrer-Williams-Ishigami-2008}   the dc conductivity 
 in the Boltzmann regime is observed to be linear in the electron (hole) density  (see fig.~\ref{f:sketch}).  This 
results in the characteristic V-shape 
 in the conductivity as a function of gate voltage. The linearity is less pronounced in suspended graphene 
where the concentration of 
charged impurities in particular has been reduced.\cite{Bolotin-Sikes-Jiang-Fudenberg-Hone-Kim-Stormer-2008,
Du-Li-Barker-Andrei-2008}
 Theoretically the linear behavior  is well described by the ordinary Drude conductivity $4\,\sigma_0=4 \,\tfrac{e^2}{2h} \ell k_\fermi $  derived from a Boltzmann equation with four degenerate (valley and real spin) incoherent bands,  provided that screened charged impurities are assumed.\cite{Nomura-MacDonald-2006,Ando-2006, Nomura-MacDonald-2007} Point-like 
 impurities on the contrary lead to a Drude conductivity that is independent of charge carrier density and therefore this model fails even on a qualitative level.\cite{Shon-Ando-1998} As  one expects the charge carrier to be mainly of one type (either electrons or holes), 
the  pseudospin band index can be left aside. 
  With interband-coherences  neglected the collision integral contains only transition rates between energy eigenstates. Such rates are simply derived with Fermi's Golden rule. The specific properties of massless Dirac electrons enter  merely into 
the transition rates as a spin-overlap factor due  to the chirality of the eigenstates  and as a Fermi momentum dependent (and therefore electron density dependent) Thomas-Fermi momentum $p_\mathrm{TF}$  due to the linearity of the spectrum.  (In a  2DEG with quadratic dispersion the screening length $\hbar/ p_\mathrm{TF}$ is in contrast independent of the Fermi momentum.)

In the Boltzmann regime $\ell k_\fermi\gg 1$ the electron-hole coherent features of Dirac electrons only manifest themselves
 if one goes beyond the lowest-order result $\sigma_0\propto \ell k_\fermi$ to look for quantum corrections. A Boltzmann approach to these quantum corrections  requires a kinetic equation that is  quantum coherent  in band indices,  for graphene the pseudospin index.\cite{Auslender-Katsnelson-2007, Trushin-Schliemann-2007, Culcer-Winkler-2007b, Liu-Lei-Horing-2008}  Interband coherent collision integrals are beyond the application range of Fermi's Golden rule. To access the ``transition rates" involving the interband components (a.k.a. off-diagonal, (pseudo)spin-precessing or Zitterbewegung components) in the collision integral one typically has to resort to a fully quantum coherent theory and then derive a Boltzmann equation by a semiclassical expansion in the space and time variables but not in the spin, which is  to be kept quantum coherent. Common approaches  for deriving quantum coherent 
kinetic equations are density matrix  approaches and nonequilibrium Green's function approaches, both with further subdivisions. 
 The former start with a single-time density-matrix-type state variable $\rho(x_1,x_2,t_1)$ governed by a quantum Liouville equation (von Neumann equation). The latter start with a double-time  correlator  $\GL(x_1,t_1,x_2,t_2)$ governed by dynamic equations derived from \eg the Kadanoff-Baym equation or the Keldysh equation. At a later stage some approximation has to be invoked to recover single-time equations. This choice of approximation is the problem of Ansatz  which will 
be discussed later.  All the mentioned treatments of pseudospin-coherence induced quantum corrections in graphene use the density matrix approach, , except Liu~\etal\cite{Liu-Lei-Horing-2008}, who use a Green's function approach together with the Generalized Kadanoff-Baym Ansatz (GKBA). In other contexts involving spin-coherent Boltzmann approaches to transport in the presence of  spin-orbit coupling  such as 
the spin Hall effect, 
Green's functions derivations have been commonly used.\cite{Mishchenko-Shytov-Halperin-2004,Shytov-Mishchenko-Engel-Halperin-2006,Raimondi-Gorini-Schwab-Dzierzawa-2006,Culcer-Winkler-2007a,Kailasvuori-2009a} However, in these cases transport was addressed to leading order rather than quantum corrections One of the two central questions of  this paper is whether 
the two approaches are equivalent in general and in particular when addressing quantum corrections to the graphene conductivity.

The general small expansion parameter for quantum corrections in a Boltzmann regime is $(\ell k_\fermi)^{-1}$. One example is the famous result by Gorkov \etal\cite{Gorkov-Larkin-Khmelnitskii-1979} for the correction $\delta \sigma/\sigma_0=  (\pi \ell k_\fermi)^{-1} \log \omega_0 \tau$ due to weak localization. The quantum correction to the graphene Drude conductivity due to electron-hole coherence is no exception.\footnote{We find that this point has not been clearly stated in the previous related studies \cite{Auslender-Katsnelson-2007, Trushin-Schliemann-2007,Culcer-Winkler-2007b}. Ref.~\cite{Auslender-Katsnelson-2007} does not seem to discuss such a parameter. Ref.~\cite{Trushin-Schliemann-2007} introduces the parameter $\alpha= 4(\ell k_\fermi)^2$ as a  ``novel electron-hole coherence parameter".  Culcer and Winkler's  treatment \cite{Culcer-Winkler-2007b} expanding in the  transition rate $|U_{\vec{k}\vec{k\pr}}|^2$, clearly displays the origin  of the expansion parameter in the spin-coherent context.  Only in  the preceding paper \cite{Culcer-Winkler-2007a}   was it stated that this expansion is one in $\hbar/\Omega_\fermi \tau$, where $\Omega$ is the spin orbit coupling, however there in a context where $\Omega_\fermi \tau \ll \aen_\fermi \tau\sim \ell k_\fermi$. 
 }     
The second central question addressed in this work is whether the series of  quantum corrections in powers of $(\ell k_\fermi)^{-1}$ starts at the order $(\ell k_\fermi)^{-2} \sigma_0$ or  already the order $( \ell k_\fermi)^{-1}\sigma_0$. 
A correction $\delta\sigma \sim \tfrac{e^2}{h}(\ell k_\fermi)^{-1}$ to the Drude conductivity $\sigma_0 \sim \tfrac{e^2}{h} \ell k_\fermi$
 would  depend on impurity concentration and impurity strength.
For screened charged impurities in graphene it 
 would  increase with decreasing electron density $n$, which would explain the onset  of convex behavior of the conductivity as one approaches the Dirac regime. Far away from the Dirac regime the contribution becomes arbitrarily small.  
 A contribution of the lower order $\delta \sigma \sim \tfrac{e^2}{h}(\ell k_\fermi)^0$  would be independent of the impurity density and impurity strength.  At least in monolayer graphene it would  also be independent  of the electron density, which  would appear as a constant  shift of the  Drude conductivity as illustrated in fig.~\ref{f:sketch}.  Electron-hole coherent effects would then remain finite arbitrarily far away from the Dirac regime,  which appears rather counterintuitive. 
  Auslender and Katsnelson~\cite{Auslender-Katsnelson-2007} found  the leading correction to the dc conductivity to be of the order $(\ell k_\fermi)^{-1}$ within a pseudospin-coherent Boltzmann approach.  \footnote{This is 
Auslender and Katsnelson's result for small enough impurity concentrations and interaction strengths, the limit of small $\Phi (\propto 1/\ell k_\fermi)$  in eqs. (84) and (85) in ref.~\cite{Auslender-Katsnelson-2007}.} 
Trushin and Schliemann\cite{Trushin-Schliemann-2007} found a leading correction of the same order
\footnote{This is the correction discussed before eq. (19) derived from eqs. (13-14). Notice that Trushin \etal also discuss another expansion, namely the expansion in the interaction range $R$, which is not an expansion in $1/\ell k_\fermi$. That they get it to be an expansion in the density is because  $R$ is taken as a constant fitting parameter.  In contrast, in the   RPA  result  $R$ is the inverse of the Thomas-Fermi momentum, which for graphene is proportional to $k_\fermi$  (see the end of our  sec.~\ref{s:model}).  In this  case the parameter $Rk_\fermi$ in graphene is independent of $k_\fermi$ and one cannot expand in the electron density.}
although their approach was qualitatively different in that  they,      
  in contrast to Auslender {\it et al.}, discarded the principal value terms in the pseudospin coherent  collsion integral. (See also ref.~\cite{Trushin-Schliemann-2008}.) 
Both papers found the result to be ultraviolet divergent for point-like impurities.\footnote{Auslender \etal treated point-like impurities. Trushin \etal treated finite range impurities, but expanded in the range parameter $R$. The first term in this expansion corresponds to the point-like limit. This issue of  is readdressed the appendix of  the sequel ref.~\cite{Trushin-Schliemann-2008}. } 
Culcer and Winkler\cite{Culcer-Winkler-2007b} studied screened charged impurities. They too  neglected principal value terms.  They solved their spin-coherent Boltzmann equation only up to order $(\ell k_\fermi)^0$, wherefore the previously found corrections $\sim (\ell k_\fermi)^{-1}$ were out of reach.  However,  solving the Boltzmann equation  by introducing a frequency dependence  lead to a quantum correction $\delta \sigma(\omega)  \sim e^2/h$  for non-zero temperatures.  As 
this correction of order $(\ell k_\fermi)^0$ vanishes in the zero-frequency limit their result  is not inconsistent with that of  Trushin \etal or, at first sight, that of Auslender \etal~Finally, Liu, Lei and Horing~\cite{Liu-Lei-Horing-2008} also found a leading correction of order $\sim (\ell k_\fermi)^{-1}$, but different to the one of Auslender~\etal or to the one of Trushin~\etal~Like Trushin~\etal they neglected principal value terms, but in addition they also argued away an interband part of the driving term.\footnote{See below equation (7) in Liu~\etal} The remaining contributions to conductivity turn out to be concentrated to the Fermi surface; the neglected driving terms lead for point-like impurities  to ultraviolet divergences as found in the previous works. We will understand that an additional source of discrepancy origins from the discrepancy in collision integrals that we will reveal between the Green's function approach together with the GKBA, used by Liu~\etal,  and the density matrix approach, used by the other authors. 

An important contribution of the paper of Auslender \etal\cite{Auslender-Katsnelson-2007} was that they realized the importance of {\it principal value terms} and  included 
them in their collision integral for graphene. However,  they did not discuss the physical origin or necessity of these  terms  that so severely complicated their analysis.  
The subsequent papers \cite{Trushin-Schliemann-2007, Culcer-Winkler-2007a} neglected them without much comment and this seems to be the rule also in other contexts where   Boltzmann approaches are applied to  spin-orbit coupled systems,  as for example in the spin Hall effect.\cite{Mishchenko-Shytov-Halperin-2004, Shytov-Mishchenko-Engel-Halperin-2006, Raimondi-Gorini-Schwab-Dzierzawa-2006, Culcer-Winkler-2007a, Kailasvuori-2009a}     
In other fields the meaning of these terms in a Boltzmann treatment  is 
rather clear (for the context of strongly interacting fermions  see \eg Lipavsky \etal\cite{Lipavsky-Morawetz-Spicka-book}). As this does not appear to be the case in the spin-orbit context, we will in the technical sections discuss the physical origin of these terms and thus motivate why we insist on keeping  them although 
 they make the interpretation of the Boltzmann equation more elusive and also considerable increase the technical challenge of solving these integro-differential equation. A witness of this increased complexity   is the  mathematical {\it tour de force} of Auslender \etal\cite{Auslender-Katsnelson-2007}, and that treatment still applying only to  the simplifying case of point-like impurities. 

In trying to answer the two mentioned central questions the present paper has two aims. 
One aim is to (re)derive a pseudospin-coherent  Boltzmann equation  and to give a systematic derivation of the leading quantum correction by including the principal value parts, but nonetheless consider screened charged impurities which are relevant to graphene. We can  consider 
a simpler problem than that of Auslender \etal by using an iterative scheme in the spirit of Culcer \etal  rather than attempting to solve the equations to all orders in $(\ell k_\fermi)^{-1}$ in one go. To illustrate the importance of the principal value terms we also solve the same Boltzmann equations with principal value terms neglected. This problem,  in contrast,  is also easily solved non-recursively.\cite{Kailasvuori-2009a}  Since we formulate the Boltzmann equation for a spin-orbit coupling of arbitrary winding number $N$ the results  also apply 
to certain models \cite{McCann-Falko-2006,Gunea-Castro-Peres-2006,Koshino-Ando-2007,Min-MacDonald-2008} for bilayer and multilayer graphene. Quantum corrections in the bilayer case $N=2$  have 
been considered by Culcer \etal\cite{Culcer-Winkler-2008} with a similar approach as in ref.~\cite{Culcer-Winkler-2007b}.  
We will refer to $N= \pm 1 $ as the  ``monolayer case" and to $|N| >1$ as the ``multilayer case". (Actual $N$-layer graphene hamiltonians can, however, 
depending on the  stacking, be written as tensor products of lower-$N$ hamiltonians, in some cases including the monolayer $N=\pm1$ hamiltonian.\cite{Koshino-Ando-2007,Min-MacDonald-2008})
We find the following results:
\begin{enumerate}
\item With principal value terms neglected, 
 the first quantum correction is of order  
$(\ell k_\fermi)^{-1}$ both in the dc conductivity as well as in the dissipative  ac conductivity.   Higher order corrections beyond the  order $(\ell k_\fermi)^{-1}$ are absent in  the monolayer case $|N|=1$, but present to infinite order in the multilayer case $|N|>1$.
\item With principal value terms included, the leading quantum correction to the dc conductivity  is of order $(\ell k_\fermi)^0$. In particular it is independent  of the impurity concentration and of the impurity strength. For screened charged impurities at a negligible distance from the graphene plane, the correction depends only on the dimensionless parameter $k_\fermi/ k_\mathrm{TF}$  (with $\hbar k_\fermi$ the Fermi momentum and $1/k_\mathrm{TF}$ the Thomas-Fermi screening length). In the monolayer case $N=\pm1$  (with $k_\mathrm{TF}\propto k_\fermi$) the correction is also independent of $k_\fermi$, that is, of the electron density, and depends only on  natural constants and the dielectric constant. 
\item For screened charged impurities the corrections are convergent. For point-like impurities the corrections are convergent in the multilayer case $|N|\geq 2$ but require 
an ultraviolet cut-off in the monolayer case $N=\pm1$.  
\item For point-like impurities in the multilayer case $|N|> 1$ the contribution from principal value terms vanishes trivially to orders $(\ell k_\fermi)^{0}$ and $(\ell k_\fermi)^{-1}$. The leading quantum correction is then the correction of order $(\ell k_\fermi)^{-1}$ derived with principal value terms neglected.
\end{enumerate} 
The first conclusion is in qualitative agreement with Trushin \etal\cite{Trushin-Schliemann-2007}. 
The ac result does not correspond to the frequency-dependent correction $\sim (\ell k_\fermi)^0$ found by Culcer \etal\cite{Culcer-Winkler-2007b,Culcer-Winkler-2008}.  The second conclusion  appears
to disagree with Auslender  \etal\cite{Auslender-Katsnelson-2007} who present  a leading  quantum correction of order  $(\ell k_\fermi)^{-1}$.\footnote{Concerning this disagreement M. I. Katsnelson communicated to us that their 
eqs. (84-85) are not the full results and 
that the full result treated in the appendix of their paper might also contain contributions of order $\mathcal{O}(1)$ 
(see also the statement at the end of their appendix) in which case the qualitative disagreement could be removed.}  
The third conclusion is in agreement with all the previous studies. The fourth conclusion is consistent with the recent paper ref.~\cite{Trushin-etal-2010}.\footnote{Conclusion 3 for multilayers is revised in the second version of our preprint thanks to discussions with M. Trushin.  Furthermore, Trushin's  observation of the vanishing of principal value terms for point-like impurities in the bilayer case $N=2$ with a density matrix approach  prompted us in the second version of the preprint to extract   from our general analysis the consequences  for the multilayer case $|N|\geq 2$ with point-like impurities, one of them being conclusion 4.} 
We wish to stress that our  electron-density-independent contribution $\sim (\ell k_\fermi)^{0}$ to the conductivity 
has a different origin than those discussed in previous studies \cite{Trushin-Schliemann-2007, Culcer-Winkler-2007b}. 

Electron-hole coherent  quantum correction gives with the shift  $(\ell k_\fermi)^0$ one mechanism for generating a residual conductivity(see fig.~\ref{f:sketch}). Our shift could therefore play a part  in  the residual conductivity observed  in the experiments by Chen \etal\cite{Chen-Jang-Adam-Fuhrer-Williams-Ishigami-2008} on monolayer graphene. This residual conductivity---estimated to be $2.6 e^2/h$---is observed to be surprisingly constant, depending at the most weakly on impurity density, in contrast to the conductivity minimum. Our contribution to the residual conductivity is independent of the impurity density. However, can in principle depend on the dielectric environment of the graphene sample. There are experiments varying the dielectric constant for example by coating the sample with ice\cite{Chen-Jang-Ishigami-Xiao-Cullen-Williams-Fuhrer-2009}. However, here the extraction of the residual conductivity appears to  be more precarious, and   it is too early to say if there is any relation with the dielectric behavior of  our contribution.

The second aim of the paper is to compare different derivations of  collision integrals for graphene and to extend the repertoire to Green's functions approaches.
Double-time approaches require, however, the choice of an Ansatz. We will see that  different choices lead to qualitatively different  general collision integrals. Some of our results are: 
\begin{enumerate}
\item The 
standard choices of the Kadanoff-Baym  Ansatz (KBA)\cite{Kadanoff-Baym-1962} or the Generalized Kadanoff-Baym Ansatz (GKBA)\cite{Lipavsky-Spicka-Velicky-1986} do not give the same general collision integral as the single-time density matrix approaches. 
\item We propose  
an alternative Ansatz---\AA{} (``anti-ordered Ansatz")---for which the translation between density matrix and Green's functions approaches can be established. 
\item  For the symmetrized KB Ansatz (\SKBA) given by the  sum of GKBA and \AA{} the principal value terms can be completely absent.
\item For spinless electrons and to zeroth order in gradient expansion the general collision integrals of all approaches coincide. This applies even in the presence of spin-orbit coupling provided the latter is small. The difference between collision integrals enters  into 
the principal value terms, which vanish for zero spin-orbit coupling (to zeroth order in gradient expansion). The difference also enters into the delta function terms, but only in parts that capture  second and higher order effects in the spin-orbit interaction.
\item Both with and without principal value terms, the leading quantum correction to the graphene Drude conductivity  depends  generally on the approach. The difference can be quantitative (for example a factor 3 for point-like impurities in the monolayer case $|N|=1$ with principal value terms neglected when comparing the result derived with a the density matrix approach to the result derived with  a Green's function approach implementing the GKBA)
 or, when principal value terms are included, even qualitative (of different order in $(\ell k_\fermi)^{-1}$ or of opposite sign, in the case of point-like impurities in  monolayers  $|N|=1$ of different ultraviolet divergent behavior). 
 \end{enumerate}
 The exception to the last conclusion is the case of point-like impurities in a bilayer $|N|=2$. In this case the  principal value terms vanish in all approaches and the remaining delta function terms of the collision integral coincide. Therefore the density matrix approach of Ref.~\cite{Trushin-etal-2010} is not concerned by the ambiguities unravelled in the present paper.

Conclusion 4, partly also discussed in ref.~\cite{Kailasvuori-2009a}, might explain why the difference between  a density matrix approach and a Green's function approach with the the GKB Ansatz does not seem to have been an issue of debate\footnote{It can be interesting to notice that 
in the field of transport through quantum dots the dichotomy of a density matrix (master equation/ superoperator) approach versus a Green's function approach has been discussed (see Timm\cite{Timm-2008}) and the question of their equivalence has been raised. There the two approaches are somewhat complementary, the first taking the perspective of the dot that gets tunneled through wheres the second that of an electron that tunnels. Although there is no firm proof of the equivalence, the general impression seems to be that the approaches should be equivalent since they both are believed to be correct.  Therefore, to the extent that the different approaches are believed to be correct also in the context of graphene, the present paper could add some new valuable input to this dichotomy. To our knowledge the issue of Ansatz has not been raised in this context.
}, even    
in similar contexts like spin-coherent Boltzmann treatments of  the spin Hall effect,  where there is a bigger variety  of applied approaches, see refs. \cite{Mishchenko-Shytov-Halperin-2004, Shytov-Mishchenko-Engel-Halperin-2006, Raimondi-Gorini-Schwab-Dzierzawa-2006, Culcer-Winkler-2007a, Kailasvuori-2009a}. With the focus on spin and charge currents to leading order, the typical neglect of principal value terms seems acceptable.  Furthermore, the energy $\ben$ of the spin-orbit coupling is typically assumed to be small compared to the spin-diagonal part $\aen\sim p^2/2m$ of the kinetic energy to motivate an expansion in the collision integrals to first order in $\ben_\fermi/\aen_\fermi$, see \cite{Shytov-Mishchenko-Engel-Halperin-2006,  Culcer-Winkler-2007a, Kailasvuori-2009a}. The difficulty of solving the Boltzmann equation analytically increases rapidly with higher orders in $\ben_\fermi/\aen_\fermi$ and  appears  to be beyond reach without an expansion. However, in the special case  of $\aen=0$, the case of graphene, many simplifications 
occur which enable an analytic solution even for a strong spin-orbit coupling. The circumstance $\aen=0$ is what allows us to  analytically explore 
the differences in predictions  between the different approaches.

Apart from the issue of the principal value terms the common framework of approximations of the related works are adopted 
in the present paper. In particular we assume  a low impurity concentration, restrict ourselves to the lowest order in Born approximation and neglect weak (anti-)localization corrections. This framework also included two further approximations:
 \begin{enumerate}
\item Terms involving the interactions (self-energy terms) are taken only to zeroth order in gradient expansion while the free part of the kinetic equation is expanded to first order in gradients (in order to recover the driving terms).
\item The Wigner transformed density matrix of the electron $\rho(\vec{p},\vec{x},t)\sim \int \der \omega\, \GL (\vec{p},\omega,\vec{x},t) $ is identified with the Landau quasiparticle distribution function  $f(\vec{p},\vec{x},t)$. The former is the quantity  for which the kinetic equations are formally derived and in terms of which the currents are expressed, the latter the quantity which in equilibrium is given by a Fermi-Dirac distribution.
 \end{enumerate}
Although each one of these approximations seems to be inconsistent at first sight, both
are standard. They are perfectly fine for the derivation of a lowest order result like the Drude conductivity or the spin Hall current. 
These approximations break down 
as soon as one has to consider quantum corrections due to strong interactions or strong fields beyond linear response. (See refs.~\cite{Lipavsky-Morawetz-Spicka-book, Haug-Jauho-book}.) 
However, it would 
be far too ambitious at present to also  properly
 account for gradient corrections and the difference between $\rho$ and $f$ in our study. Also, such a treatment would not directly be concerned with the two mentioned purposes of our study, but rather with the global goal, namely that of  a   fully consistent and systematic Boltzmann treatment of the first quantum correction. This problem is work in progress that we hope to return  
to in the future.

In this paper we have mainly graphene in mind. However, the results apply to any setting in 2d  where the electrons are described by one or several independent Dirac cones. Our results should therefore also be relevant for some   3d topological insulators like Bi$_{1-x}$Sb$_x$,   Bi$_2$Te$_3$,  Sb$_2$Te$_3$ and  Bi$_2$Se$_3$ with the 2d surface states described by Dirac electrons \cite{Fu-Kane-Mele-2007, Moore-Balents-2007,Roy-2009,Hsieh-etal-Hasan-2008, Zhang-etal-Zhang-2009,Xia-etal-Hasan-2009}, see also ref.~\cite{Hasan-Kane-2010}. In particular, the occurrence of  a single Dirac cone removes the problem of  intervalley scattering. Thus, these systems should be a better application of  those of our calculations that deal with point-like impurities, for which we would expect our assumption of negligible  intervalley scattering to  be invalid.

The outline of the paper is the following. In sec.~\ref{s:model} we present the Wigner transformed hamiltonians that we examine 
in this paper. In sec.~\ref{s:sc} the semiclassical distribution function and the Boltzmann equation for a spin-orbit coupled system are introduced.  Derivations of collision integrals with different approaches are presented in sec.~\ref{s:dci}. In sec.~\ref{s:ci} the different general collision integrals are compared and simplified for the case of  spin-orbit coupled systems, in particular graphene. Sec.~\ref{s:firsttry}  deals with 
 the solution of the Boltzmann equation neglecting principal value terms and the resulting dc conductivity. Sec.~\ref{s:acresponse} contains the ac current generalization thereof. In sec.~\ref{s:principal}, we solve the Boltzmann equation including the principal value terms and present the resulting conductivity, both dc and ac. In the beginning of that section we also discuss the physical background of the principal value terms. At the end of that section we discuss whether finite electron-hole coherent effects far away from the Dirac point make sense. Conclusions and an outlook  are given in sec.~\ref{s:conclusion}.

\section{The model}\label{s:model}
\noindent
For a semiclassical Boltzmann description  (see \eg \cite{Haug-Jauho-book,
Zubarev-etal-book,Rammer-book-1998, Rammer-Smith-1986}
one typically uses the Wigner transformed one-particle Hamiltonian. 
For the spin-orbit coupled systems  that we set out to study the hamiltonian in absence of impurities reads 
\be{ham}
H( \vec{x},\vec{p},t)~=~\aen(k)+\vec{\sigma}\cdot\vec{\ben}(\vec{k})+e\phi (\vec{x},t) 
\ee
with $e<0$ and $\vec{k}(\vec{x},\vec{p},t)=\vec{p}-e\vec{A}(\vec{x},t)$. The energy bands in the absence of electromagnetic fields are $\erg^s_\vec{k}=\aen+s\ben$ with $s=\pm$ giving the sign of the spin along the spin quantization axis $\uv{\ben}$, \ie   $\vec{\sigma}\cdot \uv{b}|\uv{b}s\rangle=s|\uv{b}s\rangle$. We want to describe a 2d system with $\vec{x}$ and $\vec{p}$ chosen to lie in the $x,y$-plane. 
We set $\hbar=1$. 

The spinless part of the dispersion is given by $\aen$. In the spin-orbit coupled systems studied in the intrinsic spin Hall effect, this is typically the quadratic dispersion $\aen=k^2/2m^*$ and  usually consitutes 
the bigger part of the kinetic energy. For monolayer graphene, the spin-orbit coupling entirely constitutes 
the kinetic part, \ie $\aen=0$.  The spin is here the pseudospin given by the bipartite lattice. Additionally, the electrons have a valley index corresponding to the two Dirac cones, K and K', as well as the real spin index. The real spin index will be  treated as trivial in the present paper. Furthermore, we will neglect inter-valley scattering to be able to treat each Dirac cone independently and therefore deal only  with the $2\times 2$ hamiltonian \eref{ham}.  This approximation should be fine if the disorder is smooth but should brake down  if the impurity potential is  too short-ranged, like in the extreme case of point-like impurities.

 For the  Dirac point K  the spin-orbit coupling for the pseudospin is given by  $\vec{\ben}=\vf \vec{k}$ 
(\ie $
\ben=\vf k$ and  
$\uv{\ben}=\uv{k}
$) 
with $\vf\approx c/300$ being the Fermi velocity. 
We consider the more general 
isotropic spin-orbit coupling  $\vec{\ben}=\ben (k)\uv{\ben} (\theta)$ with a winding number $N$ given by $\hat{\ben}_x+i\hat{\ben}_y=e^{i\theta_0+iN\theta}$ ($\theta_0$ is a constant). 
This includes the  Dirac cone K' ($\ben=\vf k$ and $\uv{\ben}=(\cos \theta,-\sin \theta)$, i.e. $N=-1$) of monolayer graphene and the hamiltonian
\be{}
H ~=~ \frac{1}{2m^*}\mtrx{cc}{0 &  (k_x\mp ik_y)^2 \\ (k_x\pm ik_y)^2 & 0}~=~
\frac{k^2}{2m^*}\mtrx{cc}{0 & e^{\mp i2\theta} \\ e^{\pm i2\theta} & 0} 
\ee
($\ben=k^2/2m^*$ and $N=2$) studied in the context of bilayer graphene as well as similar hamiltonians studied in multilayer graphene.\cite{McCann-Falko-2006,Gunea-Castro-Peres-2006,Koshino-Ando-2007,Min-MacDonald-2008} 

The total hamiltonian $H_\textrm{tot}=H+\Hi$ includes an impurity potential $\Hi (\vec{x})=\sum_n 
\singleimp (\vec{x}-\vec{x}_n)$ of non-magnetic impurities at positions $\vec{x}_n$ eventually to be averaged over. 
We distinguish between point-like impurities $\singleimp_{\vec{k}\vec{k\pr}}=\textrm{const.}$ and screened charged impurities in 2d with
\be{Vscreened}
\singleimp_{\vec{k}\vec{k\pr}}~=~    \frac{ e^2/\kappa_\mathrm{d} }
{|\vec{k}-\vec{k\pr}|+k_\mathrm{TF} } 
\ee
with the Thomas-Fermi momentum 
 $k_\mathrm{TF}= \tfrac{2\pi e^2}{\kappa_\mathrm{d}}\DS(\erg_\fermi)$ giving the  range  $L_s=1/k_\textrm{TF}$ of the screened potential.  $\kappa_\mathrm{d}$ is the dielectric constant. For a recent review on screening in graphene with an enlightening comparison of monolayers with  bilayers and 2DEGs, see ref.~\cite{DasSarma-Adam-Hwang-Rossi-review-2010}. Here we will recollect a few facts that will be important for our later discussions. 
 
It is convenient to introduce the dimensionless parameter  $\qs:=k_\mathrm{TF}/k_\fermi$ characterizing the strength of the screening.  An unscreened Coulomb interaction corresponds to  $\qs=0$. In the opposite limit $\qs\rightarrow \infty $ the potential becomes almost angularly independent and behaves in some respects as a point-like impurity, although not in all aspects. In the monolayer case, for example,  one has the unique situation that $ \DS_\fermi\propto k_\fermi$, hence $\qs$ is independent of $k_\fermi$ , implying that for short and long  screening lengths alike  does the potential behave as an unscreened Coulomb potential in that
$ \ttr^{-1}(k_\fermi)\propto \DS_\fermi k_\fermi^{-2}\propto k_\fermi^{-1}$ and therefore $\sigma_0  \sim  \ttr (k_\fermi) \erg_\fermi \propto k_\fermi^2 \propto |n|$.  Thus, not even at strong screening  does the situation turn into  that of point-like impurities, where $ \ttr^{-1}\propto k_\fermi$ predicts a Drude conductivity independent of density. For graphene on SiO$_2$ substrates, the standard value is
\be{qsml}
\qs\approx 3.2\, .
\ee
This suggests that screening is important  ($\qs >1$) and when discussing the quantum correction $(\ell k_\fermi)^0$ we will argue that this correction for  the screened potential with the given $\qs$ is nonetheless  closer related to the results for  point-like impurities than to those of an unscreened Coulomb potential. 

In bilayers and multilayers the situation is very different. Here $\qs$ decreases with $k_\fermi$ and  consequently with the density, just like in an ordinary 2DEG, but opposite to a 3DEG. Thus, the further we move away from the Dirac point, the weaker the screening and the stronger  the effect of  the interaction. Thus,  when discussing the correction $(\ell k_\fermi)^0$ we  expect that in  the vicinity of the Dirac point the screened potential has more in common with point-like impurities where as far away from the Dirac point the potential has more in common with an unscreened Coulomb potential.

\section{Semiclassical description of a spin-orbit coupled system}\label{s:sc}
\noindent
In a Boltzmann picture the spatial degrees of freedom can be  treated semiclassically.  The treatment of the peudospin must, on the contrary, remain quantum mechanical in order to capture electron-hole coherence effects.  The state of the system is described 
by the Wigner transform $\rho_{\sigma\sigma\pr}(\vec{x},\vec{p},t)$  of the  time-diagonal density matrix 
\be{}
\rho_{\sigma\sigma\pr}(x_1,x_2,t_1)~=~\langle \Psi^\dagger_{\sigma\pr}(\vec{x_2},t_1)\Psi_\sigma (\vec{x_1},t_1)\rangle= \GL_{\sigma\sigma\pr}(x_1,t_1, x_2,t_2)_{t_2=t_1}
\ee
  for electrons with spin indices $\sigma=\uparrow_z,\downarrow_z$. 
(See \eg refs.~\cite{Zubarev-etal-book, Rammer-book-1998, Rammer-Smith-1986,Haug-Jauho-book}.)
In the absence of interactions one can derive a Boltzmann equation for $\rho$  by applying the Heisenberg equation of motion on $\Psi(t_1)$, then identifying $t_2=t_1$, Wigner transforming the result, and gradient expanding to first order. The approximation to stop at first order in gradient expansion is the  {\it semiclassical} approximation,  which assumes  that the external perturbations, such as electromagnetic potentials, 
change negligibly on length and time scales of the de Broglie wavelength $\lambda_\mathrm{B}$ and the time $ \tau_\mathrm{B}=\lambda_\mathrm{B}/\vf$. Upon Wigner transformation $ X(\vec{x}_1,t_1,\vec{x}_2,t_2)\rightarrow X(\vec{x},\vec{p},t,\omega)$ (for the  time-diagonal  approaches the Wigner transformation only involves the spatial part, not time) the convolution product is translated into
$
XY\rightarrow X e^{\frac{i}{2}\grex}Y$ where the matrix product in spin remains but the convolution of time and real space variables is replaced by a Moyal product with a Poisson-bracket-like gradient $\grex$.
In a gauge invariant treatment, valid when the electromagnetic fields are weak and vary slowly
(see \eg ref.~\onlinecite{Zubarev-etal-book}), one introduces $\vec{k}(\vec{p},\vec{x},t)=\vec{p}-e\vec{A}$ and $\varomega (\omega,\vec{x},t)=\omega-e\phi$ and lets $\{\vec{x},\vec{k},t,\varomega\}$ become the new set of independent variables (\ie $\partial_{x_i}\vec{k}=0$). This  changes the gradient into
\be{gigrad}
\grex~=~
\overleftarrow{\partial}_{x_i}\overrightarrow{\partial}_{k_i} -
\overleftarrow{\partial}_{k_i}\overrightarrow{\partial}_{x_i} +
\overleftarrow{\partial}_\varomega\overrightarrow{\partial}_t-
\overleftarrow{\partial}_t\overrightarrow{\partial}_\varomega+
E_i (\overleftarrow{\partial}_\varomega\overrightarrow{\partial}_{k_i} -
\overleftarrow{\partial}_{k_i}\overrightarrow{\partial}_\varomega)+
\epsilon_{ijl}B_i \overleftarrow{\partial}_{k_j}\overrightarrow{\partial}_{k_l}\, 
\ee
with $X\overleftarrow{\partial}Y := (\partial X)Y$ and $X\overrightarrow{\partial}Y := X(\partial Y)$.

From the matrix elements of the distribution function $\rho$, one extracts the densities and current densities of charge and spin. The matrix elements are conveniently expressed in the   
 decomposition $ \rho=  \rho_0 +\vec{\sigma}\cdot\vec{\rho}$ where $\vec{\sigma}=(\sigma_x,\sigma_y,\sigma_z)$ is the vector of Pauli matrices.
Furthermore, we find it convenient to decompose the vector $
\vec{\rho}=\rho_\uv{\ben}\uv{\ben}+\rho_\uv{\cen}\uv{\cen}+\rho_z\uv{z}
$
 in its components along the basis vectors $\uv{\ben} (\theta)$, $\uv{z}$ and $\uv{\cen}(\theta)=\uv{z}\times\uv{\ben}(\theta)$ with $\partial_\theta \uv{\ben}=N\uv{\cen}$ 
,  analagous 
to the cylindrical basis vectors $\uv{k}(\theta):=\vec{k}/k$, $\uv{z}$ and $\uv{\theta}(\theta):=\uv{z}\times\uv{k}(\theta)$ with $\partial_\theta \uv{k}=\uv{\theta}$. In the spin basis $\{ |\uparrow_z\rangle, |\downarrow_z \rangle\}$ one has 
 \be{}
\rho
 =\rho_0\idm+ \rho_\uv{\ben}\uv{\ben}\cdot \vec{\sigma}+\rho_\uv{\cen}\uv{\cen}\cdot \vec{\sigma}+\rho_z\uv{z}\cdot \vec{\sigma}=
 \mtrx{cc}{\rho_0+\rho_z  & (\rho_\uv{\ben}-i\rho_\uv{\cen}) e^{-iN\theta} 
 \\
  (\rho_\uv{\ben}+i\rho_\uv{\cen}) e^{iN\theta}  &\rho_0 - \rho_z  }\, .
 \ee
 The charge density $e n$ and current density $e\vec{j}$  in phase space are derived from $en=\tr (\rho \partial H/\partial \phi) $ and $e\vec{j}=-\tr (\rho\partial H/ \partial \vec{A})$, which yields   
\be{curdef}
\begin{array}{rcccccl}
n(\vec{x},\vec{k},t) &:=&  \tr \rho & = & 2\rho_0 & 
= &  n^+ + n^- , \\
 j_i(\vec{x},\vec{k},t)&:=&\tr (\vkm_i \rho) & = & 2 \rho_0\partial_{k_i}\aen+2\vec{\rho}\cdot \partial_{k_i}\vec{\ben}
  & = & n^+ v_i^+ + n^-v_i^-+\frac{2N\ben}{k}  \rho_\uv{\cen}\hat{\theta}_i
\end{array}
\ee
with $i=x,y$. 
Here we introduced the velocity matrices $\vkm_i:=\partial_{k_i}H=\partial_{k_i}\aen +\vec{\sigma}\cdot   \partial_{k_i}\vec{\ben}$.  The spin-independent part of the velocity is $\partial_\vec{k}\aen=:\vecvo$. The band velocities are $\vec{v}^s:=\partial_\vec{k}\erg^s=\langle\uv{b}s|\vkm|\uv{b}s\rangle=v^s \uv{k}$.  The {\it intra}-band elements 
\be{}
n^\pm:=\langle\uv{b}\pm|\rho|\uv{b}\pm\rangle=\rho_0\pm \rho_\uv{b}
\ee
 give the density in each spin band $s=\pm$. The {\it inter}-band elements $\langle\uv{b}\pm|\rho|\uv{b}\mp\rangle=\rho_z\pm i \rho_\uv{\cen}$ are important for the coherent treatment of spin.  These are the elements that  oscillate in the occurrence of spin-precession. In the case of spin-orbit coupling  the imaginary component  $\rho_\uv{\cen}$ appears in the last term of  the current \eref{curdef} \footnote{Note also that this term is equally shared between the two bands;  $\langle\uv{\ben}s|\frac{1}{2} \left \{\vkm, \rho\right \} |\uv{\ben}s\rangle = n^s \vec{v}^s+k^{-1}\ben\rho_\uv{\cen} \uv{\theta} $. }. For the density matrix of a single electron this term would  contain  the oscillatory  Zitterbewegung  motion of the free  spin-orbit coupled electron. In the statistical description given by the distribution function this oscillatory motion of the free particle states average to zero over time and is therefore absent in the equilibrium distribution function ($\rho_\uv{\cen}^\textrm{eq}=\rho_z^\textrm{eq}=0$).

Throughout the paper we write the Boltzmann equation and the distribution function in the momentum-independent spin basis  $\{ |\uparrow_z\rangle, |\downarrow\rangle\}$. Some studies (\eg refs.~\cite{Auslender-Katsnelson-2007, Trushin-Schliemann-2007}) prefer to transform the Boltzmann equation  into the momentum-dependent eigenbasis
 $\{|\uv{\ben}+\rangle, |\uv{\ben}-\rangle\} $. For example, the velocity matrix for graphene ($\aen=0$)
 \be{}
\vkm=
\sigma_i (\uv{k}\partial_k+\tfrac{1}{k}\uv{\theta}\partial_\theta)\ben_i=  \uv{k} \,\vec{\sigma}\cdot\uv{\ben} \partial_k \ben + \uv{\theta}\,  \vec{\sigma}\cdot  \uv{\cen} \frac{N\ben}{k}=
\mtrx{cc} {
\uv{k}\partial_k\ben  &  i\uv{\theta}\frac{N\ben}{k}
\\
- i\uv{\theta}\frac{N\ben}{k} &- \uv{k}\partial_k\ben  
 }^\textrm{ch}   
\ee   
when written in the latter basis   (the superscript ``ch"  indicating the chirality/helicity basis). For the distribution function the relation is 
\be{}
\rho= \rho_0\idm + \rho_\uv{\ben}\uv{\ben}\cdot  \vec{\sigma} + \rho_\uv{\cen}\uv{\cen}\cdot  \vec{\sigma} + \rho_z\uv{z}\cdot \vec{\sigma} =
\mtrx{cc}{
\rho_0+\rho_\uv{\ben} & \rho_z+i\rho_\uv{\cen}
 \\ 
\rho_z-i\rho_\uv{\cen} & \rho_0-\rho_\uv{\ben}
}^\textrm{ch} =\rho_0 \idm +\rho_\uv{\ben} \sigma_z^\textrm{ch}-\rho_\uv{\cen}\sigma_y^\textrm{ch} +\rho_z \sigma_x^\textrm{ch}\, .   
\ee
The  Pauli matrices $\sigma_i^\textrm{ch}$ have non-zero momentum derivatives ($\partial_\theta \sigma_z^\textrm{ch}=-N \sigma_y^\textrm{ch}$, $\partial_\theta \sigma_y^\textrm{ch}=N \sigma_z^\textrm{ch}$ and $\partial_\theta \sigma_x^\textrm{ch}=0$).       
The eigenbasis has the attractive feature of allowing one to identify the intra-band densities  $n^\pm=\rho_0\pm \rho_\uv{\ben}$ as the diagonal matrix elements.  However, this would not be of any advantage in our treatment of graphene, where we only need  the spin  components and need to distinguish between $\rho_\uv{\ben}$, $\rho_\uv{\cen}$ and  $\rho_z$  rather  than to  view the problem  
in terms of  $\rho_{+-}(=\rho_{-+}^*)$, $\rho_{++}(=n^+)$ and $\rho_{--}(=n^-)$.  The charge component $\rho_0$  will in our problem be redundant and completely decoupled.

The spin density, \ie the polarization, is given by $s^\mu=\frac{1}{2}\tr (\sigma_\mu f)=\rho_\mu$ (with $\mu=x,y,z$).  There is not a unique way to define the spin current because  in presence of spin-orbit coupling, the real space spin polarization is not a conserved quantity. When band velocities coincide ($\vec{v}^s=\vecvo$) it is $\vec{j}^\mu =  \rho_\mu \vecvo $. A common definition is
\be{spincurr}
j_i^\mu = \frac{1}{4}\tr \sigma_\mu \{ \vkm_i, \rho\} =\rho_\mu \partial_{k_i} \aen  +\rho_0 \partial_{k_i}\ben_\mu
%=
%\hat{b}_\mu  \frac{1}{2}(n^+v^+_i - n^-v^-_i)+\frac{N\ben }{k}f_0 \hat{\theta}_i \hat{\cen}_\mu +(f_\uv{c}\hat{c}_\mu+f_z \hat{z}_\mu)v_{0i} 
\ee  
(with $\{A,B\}=AB+BA$). Notice that in the graphene case $\aen=0$ one  is confronted with the peculiar situation that the charge current is only determined by the spin density, whereas the spin current is only determined by the charge density. Here one should remember that it is electrons and {\it holes} that contribute additively to the current. The current is therefore 
  determined by the charge carrier density, that is by  the sum  of the densities of electrons  ($n^+$) and holes ($1-n^-$)---\ie by $n^+ +  (1-n^-)=1+2\rho_\uv{\ben}$---rather than by the difference $n^+-(1-n^-)=-1+2 \rho_0$. In the presence of a magnetic field one needs to involve $\rho_0$. In the Hall component of the conductivity it is the difference of electrons and holes that contributes.   

The Boltzmann equation in matrix form is given by
\be{lhs0}
i[H,\rho]+\partial_t \rho+\frac{1}{2}\{{\vkm}_i, \partial_{x_i} \rho\}+eE_i\partial_{k_i} \rho-\epsilon_{zij}eB_z \frac{1}{2} \{ {\vkm}_i, \partial_{k_j}\rho\}= \mathcal{J}[\rho]      
\ee
where the matrix-valued functional  $\mathcal{J}$ is the collision integral. 
With the  definition \eref{spincurr}  the Boltzmann equation \eref{lhs0} can  be written in  
the appealing form
\be{lhs2}
 \partial_t n +\partial_\vec{x}\cdot \vec{j}+
 e \, \partial_\vec{k}\cdot (n\vec{E}+\vec{j}\times\vec{B}) &=& 2\mathcal{J}_0,
 \nonumber 
 \\
2(\vec{s}\times \vec{\ben})^\mu +
\partial_t s^\mu + \partial_\vec{x}\cdot \vec{j}^\mu+
e\, \partial_\vec{k}\cdot (s^\mu \vec{E}+\vec{j}^\mu\times \vec{B}) &=& \mathcal{J}_\mu \, .      
\ee

Semiclassical kinetic equations deal as above with densities in phase space. 
Densities in real  space are obtained by integrating the phase space densities over momentum, \eg 
\be{}
\vec{j}(\vec{x},t)=\int \frac{\der^2 k}{(2\pi)^2} \vec{j}(\vec{x},\vec{k},t)\, .
\ee

The Boltzmann equation is typically written in  terms of the quasiparticle distribution $f_{\sigma\sigma\pr}(\vec{x},\vec{p},t)$ rather than in terms of the Wigner transformed density matrix $\rho_{\sigma\sigma\pr}(\vec{x},\vec{p},t)$. The electron distribution $\rho$ is the quantity  in terms of 
which the current and densities are defined. The quasiparticles described by $f$ on the other hand represent the free particles that satisfy the Pauli principle.  Thanks to this the equilibrium state can for $f$ be expressed simply in terms of the Fermi-Dirac distribution.  This
is of practical relevance in the analytical 
solving of  the Boltzmann equation by linearizing around equilibrium. In this paper we neglect this difference.
Thus, all expressions including $\rho$  are assumed to apply for  $f$.

\section{Derivation of collision integrals}\label{s:dci}
\noindent
The presence of, for example, electron-electron interactions, phonons or impurities, is captured in the collision integral $\mathcal{J}$. It is assumed that one is in the {\it kinetic regime}, where the de Broglie wavelength $\lambda_\mathrm{B}=1/k_\mathrm{F}$ is much shorter than the scattering length $\ell$, which translates into $\ell k_\fermi \ll 1$. We will only deal with averaged non-magnetic impurities.  The averaging over of impurity positions in $\Hi_{kk\pr}=\sum_n e^{-i(k-k\pr )x_n}\singleimp_{kk\pr}$ restores the translational invariance.\footnote{The averaging implies $ {\Hi}_{q} \rightarrow \nimp V_{q}\delta (q)\sim \nimp V(r=0)$ and $\ldots {\Hi}_{,q_1} \ldots  {\Hi}_{,q_2} \dots  \rightarrow (\ldots V_{q_1}\ldots V_{q_2}\ldots ) [\nimp^2 \delta (q_1) \delta(q_2)   +\nimp \delta(q_1+q_2)]$.
For the last term we neglect the $\nimp^2$ through the assumption of low impurity concentration.
}

In this section we will go  through 
some of the possible approaches for the derivation of semiclassical kinetic equations including collision integrals from quantum theory. One group of approaches deals directly with the density matrix $\rho(x_1,x_2,t)$ and starts from the von Neumann (quantum Liouville) equation $i\partial_t\rho =[H,\rho]$.
The second type of  approaches  
are Green's function techniques  which have 
the Kadanoff-Baym or Keldysh equations for the double-time correlator  $\GL(x_1,t_1,x_2,t_2)$ as their starting points. To recover a time-diagonal kinetic equation in the latter requires some approximation. This is the problem of  Ansatz. We will see several different candidates and therefore several different collision integrals. One of them is identical to the collision integral derived with the density matrix approaches.

\subsection{Iterative solution of the von Neumann equation (quantum Liouville equation)}
\noindent
The simplest derivation of a collision integral is probably one in which the von Neumann equation is iterated to second order in the interaction. Let $\Htot=\Ho+\Hi$ where $\Hi$ is  an  
interaction switched on at a time $t_0$ in the remote past.  The von Neumann equation  in  the interaction picture is
\be{Lioint}
i\partial_t \rho^\textrm{I}=[\Hi\INT (t),\rho^\textrm{I}]
\ee
 with 
$A\INT(t)~=~e^{i\Ho t}A(0) e^{-i\Ho t}=\evu_0^\dagger(t)A(0)\evu_0(t) $. This is easily integrated to  give
\be{intLioint}
\rhoI(t)~=~\rhoI(t_0)-i\int_{t_0}^t \der t\pr [\Hi\INT(t\pr),\rhoI (t\pr)]  
\ee
which, when inserted back into  \eref{Lioint}, yields 
\be{}
\partial_t \rho^\textrm{I}&=&-i[\Hi\INT  (t),\rho^\textrm{I}(t_0)]  -\int_{t_0}^t \der t\pr [\Hi\INT(t), [\Hi\INT(t\pr),\rhoI (t\pr)]]~ =
\nonumber \\
&=&-i[\Hi\INT  (t),\rho^\textrm{I}(t_0)]  -\int_{t_0}^t \der t\pr [\Hi\INT(t), [\Hi\INT(t\pr),\rhoI (t)]] 
-i\int_{t_0}^t \int_{t\pr} ^{t} \der t\pr [\Hi\INT(t), [\Hi\INT(t\pr), [\Hi\INT(t^{\prime\prime}),\rhoI (t^{\prime\prime})]] 
\ee
after a first and second iteration, respectively.
 So far the equations are exact.  The Born approximation allows us to get  a closed equation for $\rho$ at time $t$ to second order in the interaction $\Hi$. To this end we remove the last term in the second row. Alternatively, in the last term of the  first row replace the full evolution  with the  free evolution, \ie  let  
$\rhoI(t\pr)=\rhoI(t)$, which   
has the appearance of a Markov approximation.
Back in the \sch{}picture the assumption of free evolution reads
\be{correctmarkov}
\rho(t\pr )~=~ \evu_0(t\pr,t)\, \rho(t)\,\evu_0^\dagger (t\pr,t)~=~e^{ -i \Ho (t\pr,t)}\, \rho (t)\, e^{i \Ho (t\pr ,t)} \, 
\ee
 and after 
reorganizing evolution operators one obtains the kinetic equation in the \sch picture,
 \be{} 
\partial_t\rho (t) +  i[\Ho,\rho(t) ]= 
-i[\Hi, e^{-i\Ho (t-t_0)}\rho(t_0)e^{i\Ho(t-t_0)} ] -
  \int_0^{t-t_0}  \der \tau [\Hi,[e^{-i\Ho \tau }\Hi e^{i\Ho\tau},\rho(t)]   ] .
   \ee
 So far, this kinetic equation is locally time reversible (globally not, since we switched on the interaction).
To capture the decoherence due to other processes (phonons etc.) we do not want the state to depend on correlations in the remote past. Therefore we include a factor $e^{-\eta\tau}$ in the integral to impose this loss of memory. This introduces time irreversibility. This factor also regularizes the integral and allows us to send $t_0\rightarrow -\infty$.  In translating   
the evolution operators 
into Green's functions (see appendix~\ref{s:evogreen}), the last term can be written  as
\be{jme}
\JC[\rho]&=&-\int \frac{\der \omega}{2\pi}[\Hi,[\GoR\Hi\GoA,\rho]],
\ee
where we anticipated that the last term will become the collision integral $\JC$.

Another source of irreversibility comes with the impurity averaging procedure, a coarse graining that also captures the decoherence due to  
phonons, for example.  Then $\Hi$ in the first term becomes just a number $\sim V(r=0) $  and the commutator  vanishes.\footnote{
In the derivation of master equations for quantum dots this term 
vanishes for another reason, see \eg ref.~\cite{Timm-2008}. For $t\leq t_0$ the state is a simple product state $\rho=\rho_\textrm{dot}\otimes \rho^\textrm{eq}_\textrm{lead}$ with the leads assumed to be in equilibirum. Since $ \rho^\textrm{eq}_\textrm{lead}$ has a definite particle number whereas the hopping interaction $\Hi$ changes the particle number, tracing over the leads kills this term. 
}   
For the terms linear in $\nimp$ one finds for example (summation over repeated indices implicit) 
\be{}
(\Hi \GoR \Hi \GoA \rho)_{k k\pr } 
\longrightarrow
\nimp\delta (k-k_1+k_1-k_2) \singleimp_{kk_1}\GoR_{k_1} \singleimp_{k_1k_2}\GoA_{k_2} \rho_{k_2k\pr}= \SR_{k}\GoA_k\rho_{kk\pr},
\ee
where we introduced the retarded self-energy $\SR_k=\nimp(\singleimp \GoR\singleimp)_k$. 
However, in a term like 
\be{}
(\Hi \rho \GoR \Hi \GoA)_{kk\pr}
\longrightarrow 
\nimp
\delta (k-k_1+k_2-k\pr) \singleimp_{kk_1} \rho_{k_1k_2}\GoR_{k_2}\singleimp_{k_2k\pr}\GoA_{k\pr} 
\ee  
 the delta function seems to offer no simplification at all.

At this point we can attain further simplification if we say that in the collision integral we are not interested in any contributions which have to do with non-diagonality in momentum.  This actually amounts to saying that we are not interested in any gradient expansion corrections to the collision integral: 
\be{}
(\Hi \rho \GoR \Hi \GoA)_{kk} 
\longrightarrow \nimp  \singleimp_{kk\pr} \rho_{k\pr} \GoR_{k\pr}\singleimp_{k\pr k} \GoA_k\, .
\ee
The collision integral can then be written as 
\be{}
\JC[\rho(\vec{k},\vec{x},t) ]   ~=~  -\ki  \Wkk \int \frac{\der \omega }{(2 \pi)^2} 
(\rho_\vec{k}\GoR_\vec{k} \GoA_\vec{k\pr}+
 \GoR_\vec{k\pr}\GoA_\vec{k} \rho_\vec{k}
 -\rho_\vec{k\pr}\GoR_\vec{k\pr}\GoA_\vec{k} -
 \GoR_\vec{k} \GoA_\vec{k\pr}\rho_\vec{k\pr}
 )
\ee
with the transition matrix $\Wkk=2\pi n_\textrm{imp}|\singleimp_{\vec{k}\vec{k\pr}} |^2$ for spinless impurities.

A quantum Liouville approach was also used by Culcer \etal\cite{Culcer-Winkler-2007b}, although along different lines. The focus in that treatment was,  from the start, only the diagonal part  $f_{k}$ (not  to be confused with our $f $ for the quasiparticle distribution) of $\rho_{kk\pr}=f_k\delta_{kk\pr}+g_{kk\pr}$, closing the door to gradient expansion corrections in the interaction terms. The analogue of the iterative solution of the Liouville equation in this section is their decomposition into two coupled equations for $f_k$ and for the purely nondiagonal part $g_{kk\pr}$, the integrated equation of the latter then being inserted into the former. Until here the approaches are equivalent. The difference comes with the Markov approximation.  
Culcer \etal~use 
$f (t\pr)\rightarrow f (t)$ as opposed to $f\INT (t\pr)\rightarrow f\INT (t)$. With $\rho(t\pr)\rightarrow \rho(t)$ we find that the evolution operators cancel  each other out in a different way so that at the end they sit around the entire inner commutator rather than only around the inner $\Hi$, 
\be{culcerj}
\JC[\rho]~=~ -\int \frac{\der \omega}{2\pi}[\Hi,\GoR [\Hi,\rho]\GoA]\, .
\ee
This is indeed the result in eq. (4b) in ref.~\cite{Culcer-Winkler-2007b}. We will see that the difference between \eref{jme} and \eref{culcerj} matters for the first quantum correction. We will also understand why it does not matter for the treatment of ref.~\cite{Culcer-Winkler-2007b}.  Their recursive analysis taken to order $(\ell k_\fermi)^0$ only requires the part that is insensitive to the differences between \eref{jme} and \eref{culcerj}.

\subsection{Nonequilibirum statistical operator approach}
\noindent
The Nonequilibrium statistical operator approach (NSO)  \cite{Zubarev-etal-book} is a full-fledged second quantized field theoretical formalism and involves  more conceptual and technical ingredients from statistical physics than the approach of the previous section. It was used in ref.~\cite{Auslender-Katsnelson-2007} for  the derivation of a pseudospin-coherent collision integral for graphene. 
Like the Green's function formalism, the NSO formalism 
allows for Wick decompositions and  therefore goes beyond the approach of the previous section for non-quadratic interactions. 
 For quadratic interactions  such as 
impurities the approaches should be equivalent.

The starting point is again the Liouville equation
 \be{quantliouville}
i\partial_t \rhohat (t)-[\Htot, \rhohat(t)]~ =~ -\eta (\rhohat(t)-\rhohatrel(t)),
\ee
where $\Htot$ in contrast to the previous section is the second quantized hamiltonian. $\eta$ is a small number.  The state variable $\rhohat$ is the {\it statistical operator}. The density matrix is obtained as the expectation value of the {\it relevant operators} $P_m$ ($m$ is a composite index). In particular, with $P_{\vec{p} \sigma, \vec{p\pr} \sigma\pr}=c^\dagger_{\vec{p\pr}\sigma\pr}c_{\vec{p}\sigma}$ we have 
\be{}
\rho_{\vec{}\vec{k\pr}}^{\sigma\sigma\pr} ~=~ \langle P_{\vec{p} \sigma, \vec{p\pr} \sigma\pr} \rangle~:=~
\tr (\rhohat P_{\vec{p} \sigma, \vec{p\pr} \sigma\pr}).
\ee 

The novel ingredient on the  right-hand side 
of \eref{quantliouville} is included  {\it ad hoc} 
to  implement the irreversibility already on a fundamental level in the derivation of kinetic equations (thus well before the impurity averaging  step in 
the previous section). This is done by introducing an auxiliary statistical operator  $\rhohatrel$---the {\it relevant distribution}---that is the statistical operator with maximal entropy (and therefore minimal quantum mechanical correlations) 
 among all  operators with same expectation values for  the relevant observables 
$P_m$  ($m$ is a composite index)
\be{}
\langle P_m \rangle_\textrm{rel}~ :=~ \tr (\rhohatrel P_m)~ =~  \tr (\rhohatrel P_m) \, .
\ee
The relevant distribution $\rhohatrel$ does not evolve according to a Liouville equation like \eref{quantliouville} but is determined by the macroscopic observable $\langle P_m\rangle_t $. It serves also as an initial condition 
$\rhohat (t)=e^{-i(t-t_i)\,L}\rhohatrel (t_i)$
where $L\rho:=[\Htot,\rho]$. This evolution is however replaced by a coarse grained one, 
\be{}
\rhohat (t)~=~\frac{1}{t-t_0}\int_{t_0}^t \der t_ie^{-i(t-t_i)L}\rhohatrel{t_i}, 
\ee
(eventually $t_0\rightarrow -\infty$) to reflect that a macroscopic system 
  forgets the microscopic details of its initial state after some microscopic time. After some more steps for which we refer to the books by Zubarev \etal \cite{Zubarev-etal-book} one arrives at a general kinetic equation, here to second order in the interaction,
\be{jnso}
\frac{\partial \langle P_m\rangle }{\partial t} +\ldots~ =~\JC_m (t)~=~\JC_m^{(1)}(t)+\JC_m^{(2)}(t)+\ordo{V^3} \, .
\ee
The  first  term of order $ \Hi^1$ is  the mean-field contribution  
\be{jnso1}
\JC_m^{(1)}(t)~=~- i\langle [P_m, \Hi] \rangle^t_\rel~=~-\tr \left( \rho_\rel (t) [P_m,\Hi] \right )\, .
\ee
The  second term   is of order $\Hi^2$,
\be{jnso2}
 \JC_m^{(2)}(t)~:=~
 -\int_{-\infty}^t \der t\pr e^{\eta (t\pr-t )} \tr 
 \left (   \rho_\rel  (t\pr)\left  [\Hi, \evu_0^\dagger(t,t\pr) [\Hi, P_m]\evu_0(t,t\pr)+i\sum_n  P_n
  \frac{
  \delta J_m^{(1)}(t,t\pr)   
  } { 
      \delta \langle P_n\rangle^{t\pr} 
  } \right]  \right).   
\ee 
The first term is the familiar double commutator with the interaction. Notice that the average  is taken with respect to the relevant statistical distribution, which allows for Wick decomposition (see \cite{Zubarev-etal-book}). 
For a two-body interaction one will find that the second half, the term with $\JC^{(1)}$, does not contribute to the collision integral but cancels some anomalous terms from the first half. 
For impurities, $\JC^{(1)}$ vanishes after impurity averaging. 
Together with Born approximation (Markov approximation)  
\be{rhoevo}
\rhohatrel(t\pr) ~=~ \evu_0(t\pr ,t)\, \rhohatrel(t\pr) \,\evu_0^\dagger (t\pr ,t) +\ordo{V} \,   
\ee
which replaces the full evolution by the free evolution one arrives at  [$\evu_0$, shorthand for $ \evu_0 (t,t\pr)$]
\be{jcnso}
\JC_m^{(2)}(t)[\rho] &=& - \int_{-\infty}^t \der t\pr \, e^{\eta (t\pr-t)}
 \tr \left( 
 \evu_0^\dagger \rho_\rel(t)\evu_0 \left[
V, \evu_0^\dagger [V,P_m]\evu_0 \right]  \right) =
\nonumber \\
&=& - \int_{-\infty}^t \der t\pr \, e^{\eta (t\pr-t)}
 \tr \left( 
 \rho_\rel(t)\left[
\evu_0 V \evu_0^\dagger, [V,P_m] \right]  \right)=
\nonumber \\ 
&=&-
\int \frac{\der \omega }{2\pi} 
[\GoR V \GoA,[V,\rho]] \, . 
\ee
Notice that this is different from both \eqref{jme} and \eqref{culcerj}.
 We will see that when the collision integrals for graphene is written out explicitly in spin components the result \eqref{jme} and the result \eqref{jcnso} will coalesce.

We would like to mention that  ref.~\cite{Zubarev-etal-book}  also treats  
the {\it Cluster expansion method}, 
which is advantageous compared to NSO for going to higher order in the interaction (deriving the full $T$-matrix), whereas NSO has an advantage for dense quantum systems (for deriving two-body collision integrals 
including the Pauli blocking  factors). To second order in the 
electron-impurity interaction and to linear order in distribution functions both should apply equally. Nonetheless, the collision integral eq.~(4.2.92) 
of ref.~\cite{Zubarev-etal-book} is not equivalent to \eref{jme} or  \eref{jcnso} but to the result \eref{culcerj}. 
However, there is again a Markov approximation [see the section between eq. (4.2.24) and eq. (4.2.25)] where the full evolution is replaced by no evolution rather than by free evolution although one is in the \sch picture. Correcting this point one obtains instead the result \eref{jme} with the Cluster expansion approach.

\subsection{Green's function approach}
\noindent
In the Green's function approach one starts with general dynamic equations for the two-time correlator $\GL(t_1,t_2)$. However, solving such equations is generally difficult and therefore some approximation that limits the equations to the the time diagonal $t_2=t_1$ is desirable. This is also necessary if one wants to derive a Boltzmann type equation for $\rho(t_1)=\GL(t_1,t_1)$. We start with the discussion of the Ansatz and then turn to the derivation of semiclassical kinetic equations from the Kadanoff-Baym equations. However, we will see that even for a given Ansatz one can derive different collision integrals. 

\subsubsection{The problem of Ansatz}
\noindent
The first proposed Ansatz was the  Kadanoff-Baym (KB) Ansatz\cite{Kadanoff-Baym-1962}
\be{}
\GL (x,p,t,\omega)~=~\rho(x,p,t) A(x,p,t,\omega) 
\ee
with $A=i(\GR-\GA)$ being  the nonequilibrium spectral function. 
This is a slight nonequilibrium  modification of  the equilibrium result $\GL(k,\omega)= \FD{\omega} A(k,\omega)$ (the fluctuation-dissipation theorem) and therefore it is expected to be a good approximation close to equilibrium. For weak interactions one uses the quasiparticle approximation $A\approx 2\pi (\erg_k-\omega)$.

For nonequilibrium beyond linear response the KB  Ansatz fails. This was noted by Jauho and Wilkins\cite{Jauho-Wilkins-1982, Jauho-Wilkins-1983, Jauho-Wilkins-1984}
in Boltzmann treatments of transport in strong electric fields, where their results differed from those\cite{Levinson-1970} derived with density-matrix methods. A similar discrepancy was observed in the linear conductivity when comparing with Kubo formula calculations \cite{Holstein-1964}. Later Lipavsky \etal{} showed that the discrepancy could be cured with the modified Ansatz and coined the generalized Kadanoff-Baym Ansatz (GKBA) \cite{Lipavsky-Spicka-Velicky-1986} (see also \cite{Haug-Jauho-book,Lipavsky-Morawetz-Spicka-book})
 \be{gkba}
   \GL(\vec{x}_1,t_1,\vec{x}_2,t_2)& =&i \int \der^2 x_3\left (
  \GR(\vec{x}_1,t_1,\vec{x}_3,t_2)
\GL(\vec{x}_1,t_2,\vec{x}_2,t_2)-
 \GL(\vec{x}_1,t_1,\vec{x}_2,t_1) \GA(\vec{x_3},t_1,\vec{x}_2,t_2)    \right )~=
 \nonumber \\
&=& i\GR(t_1,t_2)\rho(t_2)-i\rho(t_1)\GA(t_1,t_2)) 
 \ee 
(spatial variables suppressed in the second line)
which reduces to the KB Ansatz  in equilibirum. 
Ref.~\cite{Lipavsky-Spicka-Velicky-1986} showed that  the right-hand side is the first term in an exact expansion, which makes it possible to address the range of validity of the Ansatz. The exact  expression respects the causal structure of the Kadanoff-Baym or Keldysh equations and  also fulfills  some other natural criteria. 
(See appendix~\ref{s:aa}). Semiclassical gradient expansion and electric field modifications can now be treated in a more consistent way. 

The GKB Ansatz seems to be most common alternative in applications when the KB Ansatz is considered insufficient. Interestingly, however, in general it does not give the same Boltzmann equation as the one derived with the mentioned density matrix  approaches (Liouville equation approaches). For the first quantum correction of graphene the difference matters. 

However, we do find  an Ansatz for which the kinetic equation obtained with a density matrix approach is also obtained from a Green's function approach, namely  if the  GKBA is replaced by the anti-ordered version  (AA for anti-ordered Ansatz)
\be{}
\GL(t_1,t_2)~=~i \GL(t_1,t_1)\GR(t_1,t_2) -i\GA(t_1,t_2)\GL(t_2,t_2)~=~i \rho(t_1)\GR (t_1,t_2) -i\GA(t_1,t_2) \rho(t_2)\, .
\ee
Although this Ansatz  violates  the causal retarded-lesser-advanced structure of KB equations and the Langreth-Wilkins rules\cite{Langreth-Wilkins-1972}, it can be derived in a similar way as the GKBA (see appendix~\ref{s:aa}).  The full result (including the omitted expansion terms) fulfills almost all the criteria required in ref.\cite{Lipavsky-Spicka-Velicky-1986}, in particular the causality requirement.  The average of  GKBA and the \AA{} gives a third alternative, here named the symmetrized Kadanoff-Baym Ansatz (\SKBA), 
\be{skba}
\GL(t_1,t_2) =\frac{1}{2}\left ( \rho(t_1)A (t_1,t_2) +A(t_1,t_2) \rho(t_2)\right )\, .
\ee
This Ansatz to  zeroth order in gradient expansion appears for example in ref.~\cite {Raimondi-Gorini-Schwab-Dzierzawa-2006}.  

Considering the importance that the issue of Ansatz has for spinless electrons in nonequilibrium beyond linear response, we believe that the issue should be even more important for graphene  calculations beyond linear response, at least when electron-hole coherent effects have to be taken into account. 

\subsubsection{The problem of identifying the collision integral}
\noindent
The 
generalized Kadanoff-Baym equation \cite{Kadanoff-Baym-1962,Langreth-Wilkins-1972} 
reads 
\be{gkbai}
\GL =\GR \SL \GA+(1+\GR \SR)G^{0<} (1+\SA \GA)
\ee
where all products  are to be interpreted as convolution products in real space/time and in spin variables. The retarded and advance components are  determined by  the Dyson equations $((\Go)^{-1}-\SR)\GR=1$ and $((\Go)^{-1}-\SA)\GA=1$. The self-energies are to first order  Born approximation given by 
\be{}
\Sigma^< &=&\nimp \Hi \GL \Hi,
\nonumber \\
\Sigma^{\textrm{R},\textrm{A}} &=&\nimp (\Hi+ \Hi G^{0\textrm{R},\textrm{A}}\Hi)\rightarrow \nimp \Hi G^{0\textrm{R},\textrm{A}}\Hi
\,   ,
\ee 
 where we will neglect the mean field terms $\sim \Hi^1$ as we are not interested in shifts of the total energy in this paper.

The term containing  $G^{0<}$ in \eref{gkbai} plays the role of  boundary conditions and vanishes when acting with $(\GR)^{-1}$ from the left or $(\GA)^{-1}$ from the right,  
\be{gkbe}
(\GR)^{-1} \GL & =&\SL\GA,
\nonumber \\
\GL (\GA)^{-1} &=& \GR\SL.
\ee
In particular, the difference gives the Kadanoff-Baym equation in differential form, which is a double-time precursor of the time-diagonal kinetic equations to be derived.  
For our discussion we want to write it in two ways. 
The first  equation, to be called \GO,  is
\be{kb1}
 [i\partial_t -H,\GL] ~=~ \SR\GL-\GL\SA +  \SL \GA -\GR \SL \,. 
\ee
 It identifies all the self-energy terms of order $\Hi^2$ with the collision integral. This is what we 
think should be done for a comparison with the Liouville equation based approaches of the previous sections, where all terms of order  $\Hi^2$ were identified with the collision integral. 
The second  equation, to be called \GT, is given by 
 \be{kb2} 
[i\partial_t-H-\Re \SR, \GL ]-[\SL, \Re \GR]~=~ i\{ \Im \SR, \GL\} -i\{\SL, \Im \GR\}\, .
\ee
 It is a frequently 
encountered starting point of Boltzmann treatments that consider renormalizations and other quantum corrections.\cite{Langreth-Wilkins-1972, Lipavsky-Morawetz-Spicka-book,Haug-Jauho-book, Mahan-book} 
Of the self-energy terms only those on 
the right-hand side are considered as the collision integral, those on the left-hand side, in contrast, as terms renormalizing the free drift. (The term $\Re \SR$ shifts for example the zero of energy.  Thereby it shifts horizontally the conductivity as a function of gate voltage, \eg shifts the minimum conductivity away from zero gate voltage.  In this context, see 
experiments  \cite{Tan-Zhang-Bolotin-Zhao-Adam-Hwang-DasSarma-Stormer-Kim-2007,Chen-Jang-Adam-Fuhrer-Williams-Ishigami-2008}.) For spinless electrons the commutators on the left-hand side vanish if one stops at zeroth order in gradient expansion. In this case one obtains the same collision integral to zeroth order in gradient expansion as with \eref{kb1}. In  general, however, and 
in particular the case of spin, the self-energy terms on the left-hand side contribute even  to zeroth order. 
We hope to address such renormalization corrections in future work. 
 In the present paper we are mainly interested in the alternative structures that might be obtained for a collision integral  from the right-hand side in \eref{kb2}.

Both equations \eref{kb1} and \eref{kb2} also hold for $G^{>}$. Within the  Keldysh formalism the same equations are derived for  $\GK=i(\GL-G^{>})$. 

Notice that the {\it quantum Boltzmann equation}\cite{Mahan-book},   obtained by  gradient expanding \eref{kb2} to first order,  is  a semiclassical kinetic equation in the variables $(\vec{x},\vec{p},t,\omega)$.  Integrating the resulting equation  over the frequency (independent energy) $\omega$ gives a Boltzmann equation.\footnote{One can also integrate over the absolute value of the momentum to get a Boltzmann equation in terms of the variables $(\vec{x},\uv{p},t,\omega)$. This is called the quasiclassical approach and is used for example in refs.~\cite{Shytov-Mishchenko-Engel-Halperin-2006, Raimondi-Gorini-Schwab-Dzierzawa-2006}. (Sometimes it is called the {\it first} quasiclassical approach and the semiclassical approach is instead called the quasiclassical approach.\cite{Lipavsky-Morawetz-Spicka-book})} In this sense  we solve Boltzmann equations in this paper, not quantum Boltzmann equations.

\subsubsection{Different collision integrals}
\noindent 
With two different ways of writing the Kadanoff-Baym equation and three different  kinds of Ansatz 
there are possibly six new collision integrals. One obvious question is, which of them corresponds to the collision integrals of the previous section. The second and {\it independent }  question is, which one is  appropriate for the problem of quantum corrections to the conductivity in graphene. 
 
In this paper we believe  we are 
able to present an answer to the first question. It seems clear that the pertinent collision integral is derived from \GO, that is \eref{kb1}. The question is what Ansatz to choose. Interestingly, it is not the GKB Ansatz, but the \AA{} that returns the collision integral \eref{jme}.
The GKBA would give  
\be{}
 \SR\GL-\GL\SA+\SL\GA-\GR\SL & =& 
 -i\Hi\GoR\Hi\rho\GoA-i\GoR\rho\Hi \GoA\Hi+i\Hi\GoR\rho\Hi\GoA+i\GoR\Hi\rho \GoA\Hi+\ldots=
 \nonumber \\ &=&-
i[\Hi,\GoR[\Hi,\rho]\GoA] +\ldots       
\ee 
where, in each term like $\SR\GL=\Hi \GoR\Hi \GoR\rho-\Hi\GoR\Hi\rho\GoA$, we neglected the parts that contain two retarded or two advanced Green's functions since such terms vanish  when one integrates over the frequency to obtain the collision integral,
\be{g1gkba}
\JC~=~-\int \frac{\der \omega}{2\pi} [\Hi,\GoR[\Hi,\rho]\GoA] \, .
\ee
We call this collision integral \GO wGKBA (w for with). It is clearly different from \eref{jme}. Interestingly it coincides with \eref{culcerj}. 

For the collision integrals \GT~derived from \eref{kb2} note, for example, that  with the GKBA we obtain
\be{jg2}
\frac{1}{2}\{ \SR-\SA, \GL\}-\frac{1}{2}\{\SL, \GR-\GA\} ~=~
-\frac{1}{2}[\Hi, \GoR[\Hi,\rho]\GoA] -\frac{1}{2}[\Hi,[\GoA\Hi\GoR,\rho] ]\, .
\ee
We denote this collision integral \GT wGKBA. 
It  takes the form of an average of \eref{g1gkba} and \eref{jme}, in the latter, however, with retarded and advanced Green's functions swapped. The other possible collision integrals are presented in the next section.

\section{Comparing collision integrals for graphene}\label{s:ci}
\noindent
The different possible general collision integrals discussed in last section will now be summarized.  When writing $\JC[\rho]=-\int \tfrac{\der \omega}{2\pi }(\ldots)$ the integrands  $(\ldots)$ of the various 
candidates are given by 
\be{jgeneral}
\begin{array}{ l r r    |      rc r }
 &\textrm{\GO} &  \textrm{GKBA}  &  [\Hi,\GoR [\Hi,\rho]\GoA] &  &
\\
\textrm{\ME{} \&} &  \textrm{ \GO } &  \textrm{\AA} & & &  [\Hi,[\GoR\Hi\GoA,\rho]] 
 \\
& \textrm{\GO } &  \textrm{\SKBA} & \frac{1}{2}  [\Hi,\GoR [\Hi,\rho]\GoA]&+& \frac{1}{2}  [\Hi,[\GoR\Hi\GoA,\rho]] 
 \\
&\textrm{\GT } &  \textrm{GKBA}  & \frac{1} {2}[\Hi,\GoR [\Hi,\rho]\GoA]&+& \frac{1}{2}[\Hi ,[\GoA \Hi \GoR,\rho] ] 
 \\ 
&\textrm{\GT } &  \textrm{\AA}  & \frac{1}{2}[\Hi ,\GoA[ \Hi,\rho] \GoR ] &+ & \frac{1} {2}[\Hi, [\GoR\Hi\GoA,\rho]]
 \\ 
& \textrm{\GT } &  \textrm{\SKBA} & 
\multicolumn{3}{c}{  \tfrac{1}{2}  {\small (\textrm{\GT wGKBA}+\textrm{ \GT w\AA})}  }
\\
 \textrm{NSO} &  & &   [\GoR \Hi \GoA,[\Hi,\rho]]   & &
\end{array}
\ee

We do not have to solve the Boltzmann equation  seven times since the first two cases are  sufficient 
to deduce all cases except the NSO case, which, however, 
will turn out to coincide with the  \GO w\AA/\ME~calculation. For example, the second term of \GT wGKBA~ is similar to  the \GO w\AA~result with retarded and advanced Green's functions swapped.
A closer  inspection (see appendix \ref{s:evogreen}) reveals that this swapping has no effect on  the delta function part of the collision integral, but changes the sign of the principal value part.  Therefore the delta function part of  \GT wGKBA is given by the delta function part of 
(\GO wGKBA +\GO w\AA)/2, whereas the principal value part is given by the principal value part of 
(\GO wGKBA -\GO w\AA)/2.  One can 
decompose the principal value part of \GO wGKBA in two parts $X$ and $Y$, where $X$ is the part that is invariant when one compares \GO w\AA{} with \GO wGKBA, whereas 
$Y$ is the part that changes sign. Then one can work out the principal value parts $\JCp=\pm \JCpx\pm \JCpy$ for all the above collision integrals, with the relative signs of $\JCpx$ and $\JCpy$, respectively, determined by the scheme
\be{pripscheme}
\begin{array}{l l | l l}
&  & X & Y
 \\
 \hline
\textrm{\GO}  & \textrm{GKBA} & +  & + \\

\textrm{\GO}  & \textrm{\AA} &  + & - \\ 

\textrm{\GO}  & \textrm{\SKBA} & + & 0 \\

\textrm{\GT}  & \textrm{GKBA} &  0 & + \\

\textrm{\GT}  & \textrm{\AA} & 0  &  - \\

\textrm{\GT}  & \textrm{\SKBA} &  0 & 0 

\end{array}
\ee
Note in particular that for the collision integral \GT w\SKBA~the principal value terms vanish completely.

The products in \eref{jgeneral} are still general convolutions. At this point the spin structure is left intact but the space and time variables  are Wigner transformed and the products are then gradient expanded in these variables. We now assume that gradient corrections in the interaction terms can be neglected.  We also replace the Wigner transformed density matrix  $\rho $ with the quasiparticle distribution $f$.  These are two approximations which might be incorrect  when 
calculating quantum corrections, but give the framework within  which we want to make a first step and compare with previous work.
Furthermore, we assume  non-magnetic impurities, $\singleimp_{\vec{k}\vec{k\pr}}^{\sigma\sigma\pr}=\delta_{\sigma\sigma\pr} \singleimp_{\vec{k}\vec{k\pr}}$ with  $\Wkk:=2\pi n_\textrm{imp}|\singleimp_{\vec{k}\vec{k}\pr}|^2$. After impurity averaging  the collision integrals become 

\be{jcn}
\begin{array}{lrcl}
\textrm{\GO wGKBA} & \JC[f(\vec{k},\vec{x},t) ] &=& 
-\ki
\Wkk 
\int \frac{\der \omega}{(2\pi)^2}
(\GoR_\vec{k}  \DT f \GoA_\vec{k\pr} +
\GoR_\vec{k\pr} \DT f \GoA_\vec{k} ) ,
\\
\textrm{\ME} 
&\JC[f(\vec{k},\vec{x},t) ]  & = & -\ki  \Wkk \int \frac{\der \omega }{(2 \pi)^2} 
(f_\vec{k}\GoR_\vec{k} \GoA_\vec{k\pr} -f_\vec{k\pr}\GoR_\vec{k\pr}\GoA_\vec{k} +\textrm{h.c.} ),
\\
\textrm{NSO} & \JC[f(\vec{k},\vec{x},t) ]  & = & -\ki
\Wkk 
\int \frac{\der \omega}{(2\pi)^2} (
\DT f \GoR_\vec{k\pr}  \GoA_\vec{k}+\GoR_\vec{k}\GoA_\vec{k\pr} \DT f ),
\end{array}
\ee
to second order in the interaction and to zeroth order in gradient expansion. The shorthand notations  $\int \frac{\der^2 k\pr }{(2\pi)^2} =:\ki$ and  $\DT f=f(\vec{k},\vec{x},t)-f(\vec{k\pr},\vec{x},t)$ were introduced. The retarded  Green's functions is the non-interacting one taken to lowest order in gradient expansion.  For the spin-orbit coupled case it is of the form
\be{gor}
\GoR_\vec{k}~ =~\sum_{s=\pm} \frac{\sproj_{\uv{\ben}s }}{\varomega^+-\erg^s_\vec{k}} ,
\hspace{2cm}
\sproj_{\uv{\ben}s}~:=~ 
\frac{1}{2}(\idm+\vec{\sigma}\cdot s\uv{\ben}_\uv{k})\, 
\ee  
with $\varomega^+=\varomega+i0^+$ and with $\varomega= \omega-\phi $ being the gauge invariant frequency variable.  
 In the Wigner representation one has  $X^\textrm{A}=(X^\textrm{R})^\dagger$. 

The three collision integrals in  \eref{jcn} would obviously be equivalent if the ingredients $\GoR$, $\GoA$ and $f$ 
commuted with each other, as is the case for spinless electrons. In the general 
non-commuting case, including the case of spin-orbit coupling, the collision integrals appear to be different. 
However, the different forms do not necessarily imply differing results. We will find that   to lowest order in $(\ell k_\fermi)^{-1}$ they all reproduce the Drude conductivity obtained with Fermi's Golden rule. To higher order in quantum corrections, however, an agreement is not at all obvious.    

Notice that the collision integral for non-magentic impurities should  generally  satisfy the property $\int_\vec{k} \JC[f(\vec{k},\vec{x},t)]=0$, expressing  that in real space $(\vec{x},t)$ the collisions cannot act as a source or drain of particles of a given spin state. For all collision integrals except the NSO  integral 
 this is  manifest since the collision integrals change sign under the renaming of dummy variables $\vec{k}\leftrightarrow\vec{k\pr}$.  In the case of  the NSO result this 
is not manifest at this level 
 but the explicit collision integral derived for graphene will turn out to have this property. 
  
For further comparison and for the explicit solution of the Boltzmann equation we will write the collision integrals  \eref{jcn}  explicitly 
in terms of the components $f_0$ and $\vec{f}$.   
To streamline  
the lengthy expressions 
some more shorthand notation is introduced.  $x\pr$ means that the quantity $x$ depends on primed variables such as $\vec{k\pr}$, $s\pr$ etc, whereas $x$ correspondingly depends on $\vec{k}$, $s$. For example $\sproj\pr = \frac{1}{2}(\idm+\vec{\sigma}\cdot s\pr\uv{\ben}_\vec{k\pr})$. Also, $\DT x := x-x\pr$, for example $\DT \erg= \erg_\vec{k}^s-\erg_\vec{k\pr}^{s\pr}$ and  $\DT (sb)=sb-s\pr b\pr$.   

Inserting \eref{gor} into \eref{jcn} gives a collision integral $\JC=\JCd+\JCp$ consisting of delta function terms and of principal value terms. 
The principal value part terms $\JCp$ are given by
\be{}
\begin{array}{lrcl}
\textrm{\GO wGKBA} 
&
\JCp [f]&=& - \ki \Wkk \frac{1}{2\pi}\sum_{ss\pr} \prip {\frac{1}{\DT \erg }} \left[
\frac{ss\pr \uv{b}\times\uv{b\pr}}{2}\cdot 
\left(-\DT \vec{f} +\vec{\sigma}\DT f_0 \right)
+\vec{\sigma}\cdot\frac{\DT (s\uv{b})}{2} \times \DT \vec{f} \right],
\\
\textrm{ \ME{} \& NSO}
&
\JCp[f] &=&-\ki \Wkk \frac{1}{2\pi}\sum_{ss\pr} \prip {\frac{1}{\DT \erg }} \left[
\frac{ss\pr \uv{\ben}\times\uv{\ben\pr}}{2}\cdot 
\left(+\DT \vec{f} +\vec{\sigma}\DT f_0 \right)
-\vec{\sigma}\cdot\frac{s\uv{\ben}+s\pr\uv{\ben\pr}}{2} \times (\vec{f}+\vec{f\pr}) \right] \, .
\end{array}
\ee
The delta function terms  are given by
\be{jcbkel}
\JCd_0[f]
&=& -\ksum{\vec{k\pr}} \Wkk\,\, 
\frac{1}{2}\sum_{s s\pr} \delta (\DT\erg )
\left[ 
\frac{1+s s\pr \uv{\ben}\cdot\uv{\ben \pr}  } {2} 
\DT f_0 + 
\frac{s  \uv{\ben}+s\pr \uv{\ben\pr}  } {2} 
\cdot \DT \vec{f}
  \right],
    \nonumber
    \\
\vec{\JC}^\delta[f]
&=&- \ksum{\vec{k\pr}} W_{\vec{k}\vec{k\pr}  }\,\, 
\frac{1}{2}\sum_{s s\pr} \delta (\DT\erg )
\left[ \DT \left(
\frac{1+s s\pr \Bkk  } {2} \vec{f}\right )  +
\frac{s  \uv{\ben}+s\pr \uv{\ben\pr}  } {2} 
\DT f_0
  \right] ,
\ee
with the matrix $\Bkk(\uv{k},\uv{k\pr})$ acting on $\vec{f}$  given by
\be{B}
\begin{array}{lrclcrcl}
\textrm{\GO wGKBA}
&
\Bkk
&: =& +
\uv{\ben} ( \uv{\ben\pr})^\mathrm{T}+
\uv{\ben\pr} ( \uv{\ben})^\mathrm{T}-
\uv{\ben}\cdot  \uv{\ben\pr}
&, &
\Bkk\pr
&:=&\Bkk
\\
\textrm{\ME} 
&
\Bkk
&:=& -
\uv{\ben} ( \uv{\ben\pr})^\mathrm{T}
+ \uv{\ben\pr} ( \uv{\ben})^\mathrm{T}
+\uv{\ben}\cdot  \uv{\ben\pr}
&, &
\Bkk\pr 
&:=&  \Bkk^\mathrm{T}
\\
\textrm{NSO}
&
\Bkk
&: =& 
+ \uv{\ben} ( \uv{\ben\pr})^\mathrm{T}-
\uv{\ben\pr} ( \uv{\ben})^\mathrm{T}+\uv{\ben}\cdot  \uv{\ben\pr}
&, &
\Bkk\pr
&:= &\Bkk
\\
\textrm{rest}
&
\Bkk
&: =& 
\uv{\ben\pr} ( \uv{\ben})^\mathrm{T}
&, &
\Bkk\pr
&:= &\Bkk^\mathrm{T}
\end{array}
\ee
The transpose of $\Bkk$ is here in only with respect to the spin-indices. In terms of momentum $\Bkk$ is not a matrix operator. Not that only for the approach \GO wGKBA and for the approaches included under "rest" is this matrix symmetric in spin indices ($\Bkk = \Bkk^\mathrm{T}$).   The ``rest" stands for the \GO w\SKBA{} as well as  all \GT{} collision integrals. As stated above, their delta function parts are just the sum of \GO wGKBA and \GO w\AA{}/\ME.   Taking into account the momentum dependence of the basis vectors 
\be{}
\uv{\ben\pr} &=&
\uv{\ben}\cos N\DT \theta-
\uv{\cen} \sin N\DT \theta ,
\nonumber \\
\uv{\cen\pr} &=&\uv{\cen}\cos N\DT \theta+
\uv{\ben} \sin N\DT \theta ,
\ee
one derives
 \be{brot}
 \begin{array}{|l|l|l|l|}
 \textrm{\GO wGKBA}   &  \textrm{\ME} & \textrm{NSO} & \textrm{ rest}
  \\
 \hline
 \Bkk\uv{\ben}= \uv{\ben\pr}   & \Bkk\uv{\ben}=\uv{\ben\pr} & \Bkk\uv{\ben}= \uv{\ben} \cos N \DT \theta+\uv{\cen} \sin N\DT \theta &  \Bkk\uv{\ben}= \uv{\ben\pr}  
 \\
 \Bkk\pr\uv{\ben\pr}= \uv{\ben}  & \Bkk\pr\uv{\ben\pr}=\uv{\ben} & \Bkk\pr \uv{\ben\pr}= \uv{\ben} 
 & \Bkk\pr\uv{\ben\pr}=\uv{\ben}
 \\
 \Bkk\uv{\cen}= -\uv{\cen\pr}  & \Bkk\uv{\cen}=\uv{\cen\pr} &  \Bkk \uv{\cen}= \uv{\cen} \cos N\DT \theta-\uv{\ben} \sin N\DT \theta 
& \Bkk\uv{\cen}=0
  \\
 \Bkk\pr \uv{\cen\pr}= -\uv{\cen}  & \Bkk\pr\uv{\cen\pr}=\uv{\cen} &  \Bkk\pr \uv{\cen\pr}= -\uv{\cen} 
 & \Bkk\pr\uv{\cen\pr}=0\, 
 \end{array}
 \ee 
Note the particular simplicity of the approaches \GO w\SKBA~and  \GT.

For studies  where a spin-coherent Boltzmann equation is linearized in a small $b$  ($b_\fermi\ll \erg_\fermi$) the terms $ss\pr\mathrm{B}$  do not appear.  Also, it should be safe to neglect the principal value terms if one is only interested in the response without quantum corrections.  In that case, the  delta function part of the collision integral is  the same in all formalisms.\cite{Kailasvuori-2009a}  However, for graphene ($\erg_\fermi=b_\fermi$) we need the full collision integral to calculate the quantum corrections and the approaches therefore differ.  However, a crucial simplification comes through   $\erg^\pm=\pm \ben$  due to $\aen=0 $. With $\DT \erg =s\ben-s\pr \ben\pr$ we obtain  (henceforth we write $\prip{1/x}$ as $1/x$)
\be{}
\begin{array}{lclcl}
\sum_{s s\pr} \delta (\DT \erg) &= &2(\delta (\DT \ben)+\delta (\ben+\ben\pr))  & &
\\ 
\sum_{s s\pr} s\delta (\DT \erg) = \sum_{s s\pr} s\pr \delta (\DT \erg)& =&0  & &
\\
\sum_{s s\pr} ss\pr \delta (\DT \erg) &=& 2(\delta (\DT \ben)-\delta (\ben+\ben\pr)) & &
\\
\sum_{s s\pr} \frac{1}{\DT \erg}=\sum_{s s\pr} ss\pr\frac{1}{\DT \erg}&=&0 & &
\\
\sum_{ss\pr}  s \frac{1}{\DT \erg} &=& 2(\frac{1}{\DT \ben} +\frac{1}{\ben+\ben\pr}) & \equiv & 2 \, \pripp  \\
\sum_{ss\pr}  s\pr \frac{1}{\DT \erg} &=& 2(\frac{1}{\DT \ben} -\frac{1}{\ben+\ben\pr}) &   \equiv &-  2 \, \pripm
\end{array}
\ee
and therefore half of the terms in \eref{jcbkel}  vanish, leaving
\be{jdirac}
\begin{array}{lcll}
\JCd[f]   &=& -\int_\vec{k\pr} \Wkk  & \left [ \right.
\delta (\DT \ben) \left( \cos^2 \frac{\DT \theta}{2}\DT f_0  +  \vec{\sigma} \cdot \DT ( \frac{1+B}{2}\vec{f}) \right )
+
%\nonumber
 \\
& &  &+   
\left .\delta (\ben+\ben\pr) \left( \sin^2 \frac{\DT \theta}{2}\DT f_0 +\vec{\sigma} \cdot \DT ( \frac{1-B}{2}\vec{f}) \right )
\right]
%\nonumber 
\vspace{0.3cm}
\\
\JCpx[f]
 &= & 
 -\int_\vec{k\pr} \Wkk &
 \vec{\sigma}  \cdot \,  
\frac{1}{2 \pi}  \left [ 
 \pripm \, 
%  \left ( 
% \frac{1 }{\ben+\ben\pr}-\frac{1}{\Delta \ben} 
% \right) 
 \uv{\ben\pr} 
 \times
 \vec{f}
-
\pripp \, 
% \left ( 
% \frac{1}{\ben+\ben\pr}+\frac{1}{\Delta \ben} 
% \right) 
 \uv{\ben} 
 \times
 \vec{f\pr}
 \right]
 %\nonumber 
 \\
 \JCpy[f]
 &= & 
 -\int_\vec{k\pr} \Wkk &
 \vec{\sigma}  \cdot \,
 \frac{1}{2 \pi}  \left [ 
 \pripp \, 
% \left ( 
% \frac{1}{\ben+\ben\pr}+\frac{1}{\Delta \ben} 
% \right) 
 \uv{\ben} 
 \times
 \vec{f}
-
\pripm \, 
%\left ( 
% \frac{1}{\ben+\ben\pr}-\frac{1}{\Delta \ben} 
% \right) 
 \uv{\ben\pr} 
 \times
 \vec{f\pr}
 \right]
 \end{array}
 \, 
 \ee
 Notice that the equations for the charge $f_0$ and the spin $\vec{f}$ are  completely decoupled for graphene.

 %To obtain the  principal value terms  for \GO wGKBA, let  $\kappa=-$  and for  \GO w\AA/\ME/NSO let $\kappa=+$.  The principal value terms of  the other cases can be constructed out of these. 
 The term with $\delta (\ben+\ben\pr)$ gives only a contribution from the point $k=-k\pr $, \ie $k=k\pr=0$ which we will neglect \footnote{    
M. Auslender communicated to us that in a careful NSO treatment  this term can be cancelled by an equivalent term coming from the $\JC^{(1)}$ part in $\JC^{(2)}$.} since we consider $k_\fermi\neq 0$. (This term 
 assures formally a continuity in the conductivity when  $k_\fermi=0^+\rightarrow k_\fermi=0$.)  Thus, the delta function part we consider is   
 \be{jdirac2}
\vec{\JC}^\delta [\vec{f}] ~ = ~ -\int_\vec{k\pr} \Wkk  &
\delta (\DT \ben)  \frac{1}{2} \left ( \vec{f}-\vec{f\pr}+\Bkk\vec{f}-\Bkk\pr\vec{f\pr}  \right) \, .
\ee
For the approach \GO w\SKBA and for  all \GT{}  approaches we obtain due to \eref{brot}  
  \be{jdiracrest}
\vec{\JC}^\delta [\vec{f}] & =&  -\int_\vec{k\pr} \Wkk 
\delta (\DT \ben)  \DT \left( \frac{1+\Bkk}{2}\uv{\ben} f_\uv{\ben} \right )
  -\int_\vec{k\pr} \Wkk  
\delta (\DT \ben) \frac{1}{2}\Delta \left(\uv{\cen}f_\uv{\cen}+\uv{z} f_z  \right )\, .
\ee
for the components $\vec{f}= \uv{\ben}f_\uv{\ben}+\uv{\cen} f_\uv{\cen}+\uv{z}f_z$.  In next section we will see that the matrix $(1+B)/2$  is responsible for the additional chirality-induced  spin-overlap factor $\cos^2 (N\DT \theta/2)$ occurring in the intraband  transition rates involving only the probability densities $ (\langle \uv{\ben}\pm |f|\uv{\ben}\pm\rangle=f_0\pm f_\uv{\ben} $ of energy eigenstates. This is how the Drude conductivity found with   Fermi's Golden rule builds in the suppression of backscattering in  monolayer 
graphene ($N=\pm1$). The ``transition amplitudes"  involving the off-diagonal components  $\langle \uv{\ben}\pm |f|\uv{\ben}\mp\rangle=f_z\pm i f_\uv{\cen}$ (the Zitterbewegung  components) are more elusive and beyond the reach of Fermi's Golden rule.  However,  for the \GO w\SKBA{} and all \GT{} approaches, the result \eref{jdiracrest} shows that 
the 
scattering of the off-diagonal components  becomes very simple  since it contains no angle dependent chirality factors but only a factor $\tfrac{1}{2}$ compared to ordinary spin independent scattering.

\section{Conductivity of graphene with principal value terms neglected}\label{s:firsttry}
\noindent
In this section we
calculate the electrical conductivity of graphene for non-magnetic impurities with the collision integrals $\JCd$ given in the previous section. 
We neglect the principal value part $\JCp$. 
We also assume low temperature so that $|\mu|=\erg_\fermi$. For notational compactness we henceforth neglect the charge unit $e$, allowing it to appear only in final results.

 Before we start, for means of comparison, we will  derive  the Drude conductivity per valley by considering only one band (electrons $\mu>0$ or holes $\mu<0$), in which case we can apply Fermi's Golden rule.  Let us consider electrons.  For monolayer graphene the one-band Boltzmann equation linearized in  the electric field ($f=\feq+\fee$ with $\feq=\FD {\vf k-\mu}$ for electrons) reads 
\be{}
\vec{E}\cdot \partial_\vec{k} \feq ~=~ -\int_\vec{k\pr} \delta (\vf \DT k )\Wkk \cos^2 \frac{\DT \theta}{2} \DT \fee~ =~
-\fee \underbrace{ \int_\vec{k\pr} \delta (\vf \DT k )\Wkk \cos^2\frac{\DT \theta}{2} (1-\cos \DT \theta)   }_{=:\ttr^{-1}},
\ee
where the transition probability $\Wkk\cos^2\frac{\DT \theta}{2}$ (contrast with $\Wkk$ for ordinary electrons) takes into account the chirality 
of the Dirac electrons, responsible for the suppression of back-scattering. The real space current is given by 
\be{drudecurrent}
\vec{j}=\int_\vec{k\pr} \vec{v}\fee~ =~- \int_\theta \int \frac{k\der k}{2\pi} 
\vf \uv{k} \ttr E_\uv{k} \partial_k  \FD{\vf k-\mu}~ =~ \vec{E}\sigma_0         
\ee 
where (reintroducing 
$e$ and $\hbar$ to the right)
\be{diracdrude}
\sigma_0~:=~ \frac{|\mu|{\ttr}_\fermi}{4\pi} ~ =~\frac{\ell k_\fermi}{4\pi}\rightarrow \frac{e^2}{2h}\ell k_\fermi 
\ee
with ${\ttr}_\fermi:=\ttr (k_\fermi) $, $k_\fermi\equiv |\mu|/\vf $  and the mean free path $\ell := \vf {\ttr}_\fermi$.  In \eqref{drudecurrent} we used the shorthand notation $\int_\theta:=\frac{1}{2\pi}\int d\theta $.   
The result \eref{diracdrude} is written  in such a way 
that it also includes the case of holes ($\mu<0$). 
To get the total Drude conductivity of graphene we multiply by a factor  of four for the degeneracy in valley index and real spin.

We turn to the coherent treatment of pseudo-spin and see that all approaches reproduce \eref{diracdrude} to lowest order in $(\ell k_\fermi)^{-1}$ but give different quantum corrections.
With 
\be{}
K(k,\theta,\theta\pr):=\int_0^\infty \frac{k\pr\der k\pr  }{2\pi} \,  \Wkk \delta (\ben-\ben\pr)\, ,
\ee
the compact notation 
$\vec{E}\cdot\partial_\vec{k} = E_\uv{k}\partial_k+E_\uv{\theta} \tfrac{1}{k}\partial_\theta$, the decomposition $f=f_0+\vec{\sigma}\cdot (\uv{\ben}f_\uv{\ben}+\uv{\cen} f_\uv{\cen}+\uv{z}f_z)$ 
with 
$\partial_\theta \uv{\ben}=N\uv{\cen} $ and $\partial_\theta \uv{\cen}=-N\uv{\ben} $ and using the table \eref{brot}  we find for \GO wGKBA  
 \be{bedirackel}
\vec{E}\cdot \partial_\vec{k} f_0 
&=&\JCd_0 =-\int_{\theta\pr} K \cos^2 \frac{N\DT \theta}{2} \DT f_0 
\nonumber, \\
 \vec{E}\cdot \partial_\vec{k} f_\uv{\ben}-E_\uv{\theta}\frac{N}{k}f_\uv{\cen} 
&=&\JCd_\uv{\ben}=
 -\int_{\theta\pr} K \left (
\cos^2 \frac{N\DT \theta}{2} \DT f_\uv{\ben}
-\frac{1}{2}\sin N\DT \theta  (f_\uv{\cen}+{f_\uv{\cen}}\pr) 
\right) 
\nonumber, \\
\vec{E}\cdot \partial_\vec{k} f_\uv{\cen}+E_\uv{\theta}\frac{N}{k}f_\uv{\ben} +2\ben f_z 
&=&\JCd_\uv{\cen}= -\int_{\theta\pr} K \left (
\sin^2 \frac{N\DT \theta}{2} ( f_\uv{\cen}+{f_\uv{\cen}}\pr)
-\frac{1}{2}\sin N\DT \theta  \DT f_\uv{\ben} \right)
, \nonumber \\
\vec{E}\cdot \partial_\vec{k} f_z- 2\ben f_\uv{\cen} 
&=& \JCd_z=-\int_{\theta\pr} K 
\sin^2 \frac{N\DT \theta}{2} \DT f_z \, .
\ee
The other collision integrals are obtained  with the ingredients in $\JCd_\mu$ replaced according to the following table
 \be{bediracnso}
\nonumber
\begin{array}{l|l|l|lr|l}
&\JCd_0 & \JCd_\uv{\ben} & \JCd_\uv{\cen} & & \JCd_z 
\\
\hline 
\textrm{\GO wG.}&{} \cos^2 \frac{N\DT \theta}{2} \DT f_0 
&{} 
\cos^2 \frac{N\DT \theta}{2} \DT f_\uv{\ben}
-\frac{1}{2}\sin N \DT \theta  (f_\uv{\cen}\pr+{f_\uv{\cen}} 
) 
&{}
\sin^2 \frac{N\DT \theta}{2} ( f_\uv{\cen}+{f_\uv{\cen}}\pr)
&+\frac{1}{2}\sin N\DT \theta  ( f_\uv{\ben}\pr-f_\uv{\ben}) 
&{} 
\sin^2 \frac{N\DT \theta}{2} \DT f_z 
 \\
\textrm{\ME}&{} \cos^2 \frac{N\DT \theta}{2} \DT f_0 
&{}
\cos^2 \frac{N\DT \theta}{2} \DT f_\uv{\ben}
-\frac{1}{2}\sin N\DT \theta  (  f_\uv{\cen}\pr-f_\uv{\cen}) 
&{}
\cos^2 \frac{N\DT \theta}{2} \DT f_\uv{\cen}
&+\frac{1}{2}\sin N\DT \theta  ( f_\uv{\ben}\pr -f_\uv{\ben} )
&{} 
\cos^2 \frac{N\DT \theta}{2} \DT f_z 
\\
\textrm{NSO}&{} \cos^2 \frac{N\DT \theta}{2} \DT f_0 
&{} 
\cos^2 \frac{N\DT \theta}{2} \DT f_\uv{\ben}
-\frac{1}{2}\sin N \DT \theta  ( f_\uv{\cen}\pr+{f_\uv{\cen}}) 
&{}
\cos^2 \frac{N\DT \theta}{2} \DT f_\uv{\cen}
&+\frac{1}{2}\sin N\DT \theta  ( f_\uv{\ben}\pr+{f_\uv{\ben}})
&{} 
\cos^2 \frac{N\DT \theta}{2} \DT f_z 
\\

\textrm{rest}&{} \cos^2 \frac{N\DT \theta}{2} \DT f_0 
&{} 
\cos^2 \frac{N\DT \theta}{2} \DT f_\uv{\ben}
-\frac{1}{2}\sin N \DT \theta   f_\uv{\cen}\pr 
&{}
 \tfrac{1}{2}f_\uv{\cen}-\tfrac{1}{2}\cos N\DT \theta f_\uv{\cen}\pr
&+\frac{1}{2}\sin N\DT \theta  ( f_\uv{\ben}\pr-{f_\uv{\ben}})
&{} 
\frac{1}{2}\DT f_z \, .
\end{array}
\\
\ee
All terms including the trigonometric factor $\sin \DT N \theta$ are in the assumed  case of symmetric scattering $K(\DT \theta)=K(-\DT \theta)$ actually the same in all approaches  since only the part including $f\pr$ can survive, whereas the part including $f$ vanishes trivially. 
Thus, the approaches differ only in the elements  $\JC_\uv{\cen}[f_\uv{\cen}]$ and $\JC_z[f_z]$. These, however,  will only enter the solution to order $\sim (\ell k_\fermi)^{-1}$.  We will see that the first quantum correction to the conductivity will depend on  $\JC_z[f_z]$.   Since the iterative solution in Culcer \etal{}\cite{Culcer-Winkler-2007b} was only  taken to order $(\ell k_\fermi)^0$ the choice of Markov approximation and in general the  choice of formalism would not have mattered.  

For further comparison with ref.~\cite{Culcer-Winkler-2007b}, notice that  eq.~(12b) for $P_\perp \JC[S_\parallel]$ (our $\JC_\uv{\cen}[f_\uv{\ben}]$) and eq.~(12c) for $P_\parallel \JC[S_\perp]$ (our $\JC_\uv{\ben}[f_\uv{\cen}]$) exactly match 
the \GO wGKBA collision integral above. The comparison was, however, already established 
in the earlier stage \eref{culcerj}. 
At the present stage, on the contrary, a comparison also with the NSO derived collision integral of Auslender \etal\cite{Auslender-Katsnelson-2007} is possible, see appendix~\ref{s:auskat}.

We proceed by linearizing the equations in the electric field with $f=\feq+\fee$ and by Fourier decomposing the components of $\fee_r=\sum_n e^{in\theta} \fee_{rn}$, ($r=0,\uv{\ben},\uv{\cen}, z$). In equilibrium we have 
\be{diracequi}
\feq_0\pm \feq_\uv{\ben}~=~\FD{\pm \vf k}~=~\Theta (\mu\mp \vf k)+\ordo{(k_\mathrm{B}T/\erg_\fermi)^2}, \hspace{1cm} \feq_\uv{\cen}~=~\feq_z~=~0\, .
\ee  
Since $\vec{E}\cdot \partial_\vec{k} \feq$ with 
\be{E}
\vec{E}\cdot\partial_\uv{k} \FD{\erg^\pm}~ =~
E_\uv{k}\partial_k \FD{\erg^\pm}~=~ \frac{e^{i\theta}\mathcal{E}^*+e^{-i\theta}\mathcal{E}}{2}\partial_k\FD{\erg^\pm}, 
\hspace{1cm} E_\uv{\theta}~=~
\frac{i}{2}(e^{i\theta}\mathcal{E}^*-e^{-i\theta}\mathcal{E})
\ee
(where $\mathcal{E}:=E_x+iE_y$) only contains $n=\pm 1$ Fourier components, we can right away conclude that $\fee_{rn}=0$ for $n\neq \pm 1$. 
It is enough to study the equation for the $n=1$ component because the $n=-1$ term is just the complex conjugate.
We find for the $n=1$ Fourier component of the linearized  Boltzmann equation 
\be{matrixbe}
\frac{\mathcal{E}^*}{2}
\mtrx{c}{\partial_k \feq_0 \\  
\partial_k \feq_\uv{\ben} \\
i\tfrac{N}{k} \feq_\uv{\ben} \\ 0 } 
=-
\mtrx{cccc}{
\IC^+  & 0 & 0 & 0 \\
0 & \IC^+ & +i\IC^\textrm{s} & 0 \\
0 & -i\IC^\textrm{s}  & \IC^\lambda & 2\ben \\
0 & 0 & -2b & \IC^\kappa}
\mtrx{c}{\fee_{01} \\ \fee_{\uv{\ben}1} \\ \fee_{\uv{\cen}1} \\ \fee_{z1} }\, .
 \ee
 In terms of 
integrals defined below one has   $\IC^\kappa=\IC^-$ for \GO wGKBA and $\IC^\kappa=\IC^+$ for \GO w\AA/ \ME/NSO.  For the \GO w\SKBA~and all \GT~approaches one has $\IC^\kappa=(\IC^++\IC^-)/2$. Likewise $\IC^\lambda=\tilde{\IC}^+$ for  \GO wGKBA, 
$\IC^\lambda=\IC^+$ for 
\GO w\AA/\ME/NSO and  just the average for the rest.  . 
\be{ic}
\begin{array}{rcl|cl}
 & & & & \textrm{for} \,\, N=\pm 1
  \\
 \hline
\IC^+ &:= &\int_{\theta\pr} K \cos^2 \frac{N\DT \theta}{2}(1-e^{-i\DT \theta})\equiv \ttr^{-1} &=& \frac{1}{4}\int_{\theta\pr}  K (1-\cos 2\DT \theta) 
 \\
\tilde{\IC}^+&:= & \int_{\theta\pr}  K \sin^2 \frac{N\DT \theta}{2} (1+e^{-i\DT \theta }) &=& \IC^+ \quad 
 \\
\IC^- &:=& \int_{\theta\pr}  K \sin^2 \frac{N\DT \theta}{2}
(1-e^{-i\DT \theta }) &=& \frac{1}{4}\int_{\theta\pr}  K (3-4 \cos \DT \theta+\cos 2\DT \theta) 
\\
 i \IC^\mathrm{s} &:=&- \int_{\theta\pr}  K \frac{\sin N\DT \theta}{2}  e^{-i\DT \theta }=\frac{i}{2} \int_{\theta\pr}  K \sin N\DT \theta \sin \DT \theta& \equiv & 
\pm
 i \IC ^+  
\end{array}
\ee
where we used that $K(-\DT \theta)=K(\DT \theta)$. Notice that  $\IC^\mathrm{s}$ is odd in $N$ whereas the other integrals $\IC$ are even in $N$.

The equation for $f_0$ is decoupled from the other components and is solved by $\fee_{01}=-\frac{1}{2}\mathcal{E}^* \ttr \partial_k \feq_0$ with $\ttr := (\IC^+)^{-1} $. The other components are found by inverting the remaining $3\times3$ matrix,
\be{ffromM}
 \mtrx{c}{
 \fee_{\uv{\ben}1} \\
 \fee_{\uv{\cen}1} \\
 \fee_{z1}
 }
 &=&
-\frac{\mathcal{E}^*}{2|M|}
 \mtrx{c}{
(4\ben^2 +\IC^\lambda\IC^\kappa) 
\partial_k \feq_\uv{\ben}  + 
 \IC^\textrm{s}  \IC^\kappa \tfrac{N}{k}\feq_\uv{\ben}  \\
i \IC^\kappa (\IC^\textrm{s} \partial_k \feq_\uv{\ben} + \IC^+\tfrac{N}{k}\feq_\uv{\ben}   )
\\
i  2  \ben (\IC^\textrm{s}  \partial_k \feq_\uv{\ben}+ \IC^+\tfrac{N}{k}\feq_\uv{\ben}   )
} .
\ee
with the determinant $|M|=4\ben^2 \IC^+ + \IC^+ \IC^\lambda\IC^\kappa  - (\IC^\textrm{s})^2 \IC^\kappa$. Adding up the two Fourier components $n=\pm 1$ one obtains 
\be{feedc}
\begin{array}{rclcl}
\fee_\uv{\ben} &=&  -E_\uv{k} 
\frac{
(4\ben^2 +\IC^\lambda\IC^\kappa) 
\partial_k \feq_\uv{\ben}  + 
 \IC^\textrm{s}  \IC^\kappa \tfrac{N}{k}\feq_\uv{\ben} 
}
{
4\ben^2 \IC^+ + \IC^+ \IC^\lambda\IC^\kappa  - (\IC^\textrm{s})^2 \IC^\kappa
}
&\longrightarrow&
  -E_\uv{k} \left ( \frac{1}{\IC^+}\partial_k \feq_\uv{\ben}+
\frac{\IC^\kappa}{4\ben^2}
\left(  \partial_k \feq_\uv{\ben} + \tfrac{1}{k}\feq_\uv{\ben}      \right) \right )
,  
 \\
 \fee_\uv{\cen} &
=&   -E_\uv{\theta}
\frac{
 \IC^\kappa (\IC^\textrm{s} \partial_k \feq_\uv{\ben} + \IC^+\tfrac{N}{k}\feq_\uv{\ben}   )
} 
{
4\ben^2 \IC^+ + \IC^+ \IC^\lambda\IC^\kappa  - (\IC^\textrm{s})^2 \IC^\kappa
} 
 & \longrightarrow &-E_\uv{\theta} \frac{N \IC^\kappa}{4\ben^2}
 \left(\partial_k \feq_\uv{\ben} + \tfrac{1}{k}\feq_\uv{\ben}     \right) 
 ,
  \\
  \fee_z &=&   -E_\uv{\theta}
 \frac{
 2  \ben (\IC^\textrm{s}  \partial_k \feq_\uv{\ben}+ \IC^+\tfrac{N}{k}\feq_\uv{\ben}   )
  } 
  {
4\ben^2 \IC^+ + \IC^+ \IC^\lambda\IC^\kappa  - (\IC^\textrm{s})^2 \IC^\kappa
} 
&\longrightarrow 
  & -E_\uv{\theta} \frac{N }{2\ben}
\left(  \partial_k \feq_\uv{\ben} + \tfrac{1}{k}\feq_\uv{\ben}      \right)
\end{array}
\, ,
 \ee
 The far right shows the monolayer case $N=\pm1$. The result \eref{feedc} is the solution to all  orders in $(\ell k_\fermi)^{-1}$. The expansion can be carried out by expanding the denominator in $\IC/\ben $.    
In the monolayer case $N=\pm 1$ the denominator simplifies as $|M| =4 \ben^2 \IC^+ $ due to 
 $\IC^\lambda=\IC^+$ and  $\IC^\textrm{s}\equiv N \IC^+$, therefore no expansion is possible.   Consequently, contributions beyond the leading order correction will be absent in the monolayer case. 
Our monolayer result has a structure similar to that of eqs. (13-14) in Trushin \etal~\cite{Trushin-Schliemann-2007},  however, up to the sign of the $\tfrac{1}{k}\feq_\uv{\ben}$ part. The sign difference, which originates from the sign of $E_\uv{\theta}\tfrac{N}{k}f_{\uv{\ben}}$
 in the left-hand side of  \eref{bedirackel}, will determine the sign of $\sigma^{II}$ to be introduced below.

The charge  current $\vec{j}$ in momentum space, see eq.~\eref{curdef}, is in the graphene case ($\aen=0$) given by
\be{diraccurr}
\vec{j}(\vec{k})~=~2\uv{k}f_\uv{\ben}\partial_k b +2\uv{\theta}\vec{f}_ \uv{\cen}
 \frac{N\ben}{k} ~\longrightarrow ~ 2\vf (\uv{k}f_\uv{\ben}+N\uv{\theta}f_\uv{\cen})
\ee   
with the monalayer case $N=\pm1$ to the right. 
With $\int_\theta \uv{k}E_\uv{k}=\int_\theta \uv{\theta}E_\uv{\theta}=\vec{E}/2$   one obtains the current in real space 
\be{}
\vec{j}~=~ \int_\vec{k} \vec{j}(\vec{k}) =(\sigma^I+\sigma^{II})\vec{E } \, .
\ee
The conductivity is given by  the contributions
\be{conducorrwopp}
\begin{array}{rclcl}
\sigma^I & = & -  \int \frac{k\der k}{2\pi} \,
\frac{
( 4\ben^2 + \IC^\kappa \IC^\lambda) \partial_k \ben
+\IC^\kappa \IC^\mathrm{s}\tfrac{N\ben}{k} }
{4\ben^2 \IC^+ + \IC^+ \IC^\lambda\IC^\kappa  - (\IC^\textrm{s})^2 \IC^\kappa} \partial_k \feq_\uv{\ben}
&\longrightarrow&
 -\frac{ \vf}{2\pi}   \int k\der k
\left (
\frac{1}{\IC^+} +\frac{\IC^\kappa }{2\ben^2} 
\right)  \partial_k \feq_\uv{\ben}
  \,
\\
\sigma^{II} & = & -  \int  \frac{\der k}{2\pi} \,  
N
\frac{ 
   \IC^\kappa 
 \IC^\mathrm{s}   \partial_k \ben
 +
\IC^\kappa \IC^+ \tfrac{N\ben}{k}  
   }
{
4\ben^2 \IC^+ + \IC^+ \IC^\lambda\IC^\kappa  - (\IC^\textrm{s})^2 \IC^\kappa} \feq_\uv{\ben}
&\longrightarrow&
-\frac{\vf}{2\pi} 
\int  \der k \, 
 \frac{ 
  \IC^\kappa   
   }
{
2\ben^2 
}
 \feq_\uv{\ben}
  \, ,
  \end{array}
\ee
where for the monolayer case to the far right it was used that $\partial_k\ben =\tfrac{\ben}{k}=\vf$ for all $k$. Because of $\IC^\mathrm{s}_{-N}=-\IC^\mathrm{s}_{N} $ the conductivity is invariant of the sign of $N$.

As a part of 
$\sigma^I$ we recognize the Drude contribution $\sigma_0 =-  \int \tfrac{k\der k}{2\pi}  \tfrac{\partial_k\ben }{\IC^+}
\partial_k \feq_\uv{\ben} = \ell k_\fermi/4\pi$. The contributions $\sigma^{II}$ and $\delta\sigma^I=\sigma^{I}-\sigma_0$ are quantum corrections, both of order $(\ell k_\fermi)^{-1}$ and  for $|N| > 1$ containing also higher powers of $(\ell k_\fermi)^{-1}$.  Notice that there is no contribution of the order $(\ell k_\fermi)^0$. 

At $T=0$ one  has from \eref{diracequi} that  $\feq_\uv{\ben}=-\tfrac{1}{2}\Theta (k-k_\fermi)$ and $\partial_k \feq_\uv{\ben}= -\tfrac{1}{2} \delta (k -k_\fermi)$.   The conductivity including  the leading quantum correction $\sim (\ell k_\fermi)^{-1}$  [for general $N$ obtained by a truncated expansion  of the denominator in \eref{conducorrwopp}] becomes
\be{sigmab} 
\begin{array}{rclcl}
\sigma^I & = & \sigma_0 \left(
 1 +\frac{
  \IC^\kappa_\fermi (\IC^\mathrm{s}_\fermi)^2 
  }{
  4\ben_\fermi^2 \IC^+_\fermi
  }
  +
  \frac{
  N\IC^\kappa\IC^\mathrm{s}
  }{
 4\ben_\fermi v_\fermi k_\fermi  
  } \right)
 +\mathcal{O}((\tfrac{1}{\ell k_\fermi})^{2})
  & \longrightarrow&
 \sigma_0 \left(
 1 +\frac{\IC^\kappa_\fermi  \IC^+_\fermi }{2\ben_\fermi^2}  \right),  
 \\
\sigma^{II} & = &\frac{1}{4\pi} \int_{k_\fermi}^\infty \der k 
\left(   
\frac{ N \IC^\kappa \IC^\mathrm{s}  \partial_k \ben}{4\ben^2 \IC^+}
+ 
\frac{N^2\IC^\kappa}{4\ben k }  \right )
+\mathcal{O}((\tfrac{1}{\ell k_\fermi})^{2})
& \longrightarrow  & \frac{1}{8\pi } \int_{\ben_\fermi}^\infty \der \ben \frac{\IC^\kappa}{\ben^2 } \, ,
\end{array} 
\ee 
where we write out the monolayer ($N=\pm 1$) result to the far right. With real spin and valley degeneracies included the conductivity of graphene is  $\sigma_\textrm{graphene} = 4 ( \sigma^{I}+\sigma^{II})$. 
The leading correction depends on $\IC^\kappa$ but not on $\IC^\lambda$. Thus the difference in $\IC^\kappa$---that is the one in $\JC_z[f_z]$---is the one that is crucial for the discrepancy between approaches.

The leading corrections are positive within all the approaches. When we later  include principal value terms this will no longer be the case.

 For screened charged impurities $W(k,\DT \theta)=2\pi n_\textrm{imp}(2k\sin \tfrac{|\DT \theta|}{2} +k_\textrm{TF})^{-2}$ the integral in $\sigma^{II}$ is 
convergent.  
For point-like impurities  $K(k,\Delta \theta)=\frac{k}{2\pi v}W_0 $ with $\Wkk=W_0=\textrm{const} \propto n_\textrm{imp}$   the integral for $\sigma^{II}$  has
a logarithmic divergence in the monolayer case  (since $\IC\propto k/v=k/\vf$) unless an ultraviolet cut-off is introduced. Let us nonetheless make the  observation that  $\IC^+ = K/4$ and $\IC^-=3 K/4$. Given a cut-off, the 
 leading quantum correction is larger   
 by a factor of 3 with the approach  \GO wGKBA
($\IC^\kappa=\IC^-$) compared to the approaches \GO w\AA/\ME/NSO  ($\IC^\kappa=\IC^+$). Other approaches lie midway between these two results. 
 
For point-like impurities in the multilayer case $|N|\geq 2$ all the approaches coincide because
\be{}
\IC^+=\frac{1}{2}K =\tilde{\IC}^+=\IC^- 
\ee
hence $\IC^+=\IC^\lambda=\IC^\kappa=:\IC$. Furthermore $\IC^\mathrm{s}=0$. With 
 $\ben =\alpha k^N$  and $\IC= \tfrac{1}{2}K=\tfrac{ k^{2-N} }{4\pi N \alpha}W_0$ the  $T=0$ limit of the untruncated form \eref{conducorrwopp} is easy to carry out.  The result is (here with $e$ and $\hbar$ reintroduced)
  \be{plmulti}
 \begin{array}{rcl}
 \sigma^I & =& \sigma_0 \\
 \sigma^{II} & = & \frac{e^2}{2h} \frac{N^2}{4(|N|-1)} \arctan \frac{|N|}{2\ell k_\fermi} \, .
 \end{array} 
 \ee
 All integrals converge without any ultraviolet cut-off. The $N=2$ case of   \eref{plmulti} was worked out  in collaboration with M. Trushin \etal, see  ref.~\cite{Trushin-etal-2010}.

The treatment in this section would until equation \eref{conducorrwopp} also  hold for the density matrix $\rho$, provided that $\rho_\uv{\ben}^\mathrm{eq}$ is independent of $\uv{k}$ and that $\rho_\uv{\cen}^\mathrm{eq}=\rho_z^\mathrm{eq}=0$. The difference would come in the last step \eref{sigmab}. If  the difference between $\rho_\uv{\ben}^\mathrm{eq}$ and $\feq_\uv{\ben}$  is 
of order $\Hi^r$, $r\leq 4$ it could in principle change the value or the order of the leading quantum correction.

\section{AC response}\label{s:acresponse}
\noindent
For an ac field $\vec{E}(t)=e^{i\omega t} \vec{E}$ one obtains with the Ansatz $\fee_\vec{k}(t)=e^{i\omega t} \fee_\vec{k}$ the Boltzmann equation
\be{beac}
e^{i\omega t } \left(i\omega \fee_\vec{k}+i[H,\fee_\vec{k}]+\vec{E}\cdot \partial_\vec{k}\feq_\vec{k} -\JC[\fee_\vec{k}]\right)~=~0
\ee
from which the Drude result \eref{diracdrude} is modified into 
\be{diracdrudeomega}
\sigma_0 (\omega)~:=~ \frac{v_\fermi k_\fermi }{4\pi ({\ttr}_\fermi^{-1}+i\omega)}  ~=~\frac{\ell k_\fermi}{4\pi (1+i\omega{\ttr}_\fermi )}\, .
\ee
In the coherent treatment of spin of the previous section the term $i\omega \fee e^{i\omega t}$ 
enters as a diagonal matrix $i\omega \idm_4$ in equation \eref{matrixbe}, \ie we obtain the ac result from the dc result for the $n=\pm1$ Fourier components  with the substitutions 
\be{acsub}
\IC^{+}\rightarrow \IC^{+}+i\omega, \hspace{2cm}
 \IC^{\kappa}\rightarrow \IC^{\kappa}+i\omega, \hspace{2cm}
  \IC^{\lambda}\rightarrow \IC^{\lambda}+i\omega, \hspace{2cm}
 \IC^\mathrm{s}\rightarrow \IC^\mathrm{s}\, ,
 \ee
  Since the Boltzmann equation \eref{beac} with $\omega\neq 0$ is no longer real, the $n=-1$ component of the solution is no longer obtained by simply complex conjugating  the $n=+1$ component of the solution. The correct $n=-1$ component is obtained by carrying out the substitution \eref{acsub}  {\it after} the complex conjugation. This the same as carrying out the substitution in the full dc solution where the two Fourier components have already been added up. 
  With \eref{acsub} the  dc result  \eref{feedc} is modified into the ac result   
\be{feeac}
\begin{array}{rcll}
\fee_\uv{\ben} &=&  -E_\uv{k} \frac{1}{|M|} \left( \left[4\ben^2 +( \IC^\lambda +i\omega)( \IC^\kappa+i\omega)\right] 
\partial_k \feq_\uv{\ben}  + 
 \IC^\mathrm{s} \left[\IC^\kappa+i\omega\right]  \frac{1}{k}\feq_\uv{\ben} \right)
&   \longrightarrow - E_\uv{k} \partial_k \feq_\uv{\ben} \frac{1}{i\omega} 
 , \\
 \fee_\uv{\cen} &=&  -E_\uv{\theta} \frac{1}{|M|}
 (\IC^\kappa+i\omega) \left( \IC^\mathrm{s} \partial_k \feq_\uv{\ben} +\left[\IC^+ +i\omega\right] \frac{1}{k}\feq_\uv{\ben}     \right)
 &\longrightarrow -E_\uv{\theta} \frac{1}{k} \feq_\uv{\ben} \frac{i\omega}{4\ben^2-\omega^2} 
 , \\
  \fee_z &=&  -E_\uv{\theta} \frac{1}{|M|}
 2\ben \left(  \IC^\mathrm{s}\partial_k \feq_\uv{\ben} + \left[\IC^{+}+i\omega \right]\frac{1}{k}\feq_\uv{\ben}      \right)
 & \longrightarrow -E_\uv{\theta} \frac{1}{k} \feq_\uv{\ben} \frac{2\ben}{4\ben^2-\omega^2}, \, 
 \end{array}
 \ee
 with the determinant $|M|$ also shifted according to \eref{acsub}.  We complemented  this with the pure sample limits ($\Wkk\rightarrow 0$) to the far right. 
  
 Ac  
terms are derived from dc terms by replacing a real quantity $\IC$ by an imaginary quantity $ i \omega$. Thus the real frequency-dependent contributions in $\sigma(\omega)$ step up or down in even powers of $\omega /\IC$, whereas an odd power would be  needed 
to derive a $(\ell k_\fermi)^0$ correction in $\Re \sigma(\omega)$ from the dc result $\sigma_\textrm{Drude}\sim \ell k_\fermi$ or its dc corrections $\sim (\ell k_\fermi)^{-1}$. According to this argument, there are no corrections $(\ell k_\fermi)^0$ to $\Re \sigma (\omega)$. This result should be contrasted with the frequency-dependent corrections  of order $(\ell k_\fermi)^0$ found by  Culcer \etal (eq. (31) and eq. (27)  in \cite{Culcer-Winkler-2007b}).  

For 
further details, see 
appendix~\ref{s:acdetails}.

\section{First quantum correction with principal value terms included}\label{s:principal}     
\noindent
In this section we include the principal value terms and recalculate the first quantum correction to the conductivity, this time in a recursive solution 
in the spirit of ref.~\cite{Culcer-Winkler-2007b}. The recursion is only taken to order $(\ell k_\fermi)^0$ in the distribution function, but could in principle be reiterated to access terms of order   $(\ell k_\fermi)^{-1}$ and higher as discussed in appendix~\ref{s:secondrecursion}.  
%To order $(\ell k_\fermi)^0$ the recursive result  exactly matches  
%the non-recursive solution of sec.~\ref{s:firsttry}. 
However, the structure of the recursive setting makes it clearer  why the  $(\ell k_\fermi)^0$ correction goes only into $f_z$, whereas $f_\uv{\ben}$, $f_\uv{\cen}$---and by consequence the current---get no contribution if principal value terms are neglected. When the latter are included the recursive setting  clearly shows why  $f_\uv{\ben}$ and  $f_\uv{\cen}$ then can get 
corrections already of the order  $(\ell k_\fermi)^0$.

We will start with giving  the physical reason for taking principal value terms seriously in our semiclassical kinetic equations. 
General collision integrals  derived within a quantum coherent approach 
typically contain principal value terms (a.k.a. reaction terms, off-shell terms, off-pole terms) alongside  with the delta function terms (a.k.a elastic terms, on-shell terms, pole terms). 
The delta functions convey the sharpness in energy of the idealized semiclassical quasiparticles. The quasiparticles are the almost free particles 
 that will distribute themselves according to the Fermi-Dirac distribution in equilibrium, whereas the electrons due to the interaction will be given by a distribution with fatter tails.\cite{Lipavsky-Morawetz-Spicka-book}
The principal value terms are a reminiscence of the quantum coherent nature of the underlying particles and captures the deviation from the classical point-like ``billiard ball" picture conveyed by  the  fully semiclassical (\ie quantum incoherent) Boltzmann equation. One such example is the principal value terms related to the quickly decaying  coherences coming from  the redressing of the quasiparticles within the interaction radius.\cite{Lipavsky-Morawetz-Spicka-book} The corresponding decay time (the collision time, the spent within the interaction radius) is in the kinetic regime by assumption much shorter than the relaxation time (roughly $\ttr$), wherefore the electron quickly recovers its asymptotic quasiparticle nature after one  collision on its way to the next. In  the spectral function the off-pole part is the broad background around the quasiparticle peak.\cite{Bruus-Flensberg-book} For spinless electrons there are ways of separating out the off-pole part from the quantum kinetic equation, with  the remains becoming the standard Boltzmann equation for the quasiparticles.\cite{Lipavsky-Morawetz-Spicka-book}

The electron-hole coherence (or spin-coherence), too, is   
a deviation from the fully semiclassical particle picture, in this case not because of  interaction effects but because of the Zitterbewegung due to the spin-orbit coupling. It is therefore no surprise that a spin-orbit coupling  contributes with its own principal value terms adding to  those related to the quasiparticle redressing. However,  this time we do not want to separate out the principal values in deriving a Boltzmann type equation 
since the Zitterbewegung is 
known to be inherent in the asymptotic free particle.  When we want to derive a kinetic equation  while keeping track of electron-hole coherent  effects, we should therefore  keep the corresponding principal value terms. To our knowledge this physical motivation has not been given before in the graphene context nor in related  fields (\eg spin Hall effect, anomalous Hall effect). The only reference we know of that treats principal value terms due to spin (although not spin-orbit coupling) is  the Green's function derivation  by Langreth and Wilkins\cite{Langreth-Wilkins-1972} of a Boltzmann equation for electrons interacting with localized spins.
There the  principal value terms are found to give important renormalizations.  

The technical problem with including principal value terms is that the two momenta $\vec{k}$ and $\vec{k\pr}$ in the collision integral are no longer confined to sit on the same surface.  This is in contrast with the previous situation  which allowed us to plug out the Fourier coefficients $f_{rn}(k)$ from the integrals, of which the remains become matrix elements like $\IC^\pm$ etc. (The problem of solving the Boltzmann equation to all orders in $(\ell k_\fermi)^{-1}$ 
then reduced to a matrix inversion.) With the principal value terms 
we have to confront difficult integro-differential equations. In ref.~\cite{Auslender-Katsnelson-2007} an analytical solution to all orders in $(\ell k_\fermi)^{-1}$ was 
obtained for point-like impurities.
The solution for screened charged impurities is still unknown and will be addressed here within the recursive scheme. 
It happens that  we do not run into the problem of finding unknowns inside of the integrals $\JCp$ with principal values. Therefore, no simplifying assumption about the  potential is needed. 

From the principal value terms in \eref{jdirac} we find in the case \GO w\AA /\ME{}/NSO
\be{xnso}
\JCp_0 & = & 0,
\nonumber 
\\
\JCp_\uv{\ben} & = & 
+\int_\vec{k\pr} \frac{\Wkk}{2\pi} \pripm \sin N \DT \theta f_z \pr \, ,
\nonumber 
\\
\JCp_\uv{\cen} & = & 
-\int_\vec{k\pr} \frac{\Wkk}{2 \pi} 
(\pripp-\pripm \cos N \DT \theta) (f_z+f_z\pr)  \, ,
\nonumber 
\\
\JCp_z & = & 
-\int_\vec{k\pr} \frac{ \Wkk}{2\pi}
\left\{ 
\pripm (\sin N \DT \theta f_\uv{\ben} +\cos N \DT \theta f_\uv{\cen} + f_\uv{\cen}\pr   )
+\pripp  (\sin N \DT \theta f_\uv{\ben}\pr  - \cos N \DT \theta f_\uv{\cen} \pr -  f_\uv{\cen}\pr   )
\right\}
 \, .
\ee 
For a comparison with Auslender \etal\cite{Auslender-Katsnelson-2007}, see appendix~\ref{s:auskat}. 
For obtaining all the other cases we express $\JCp$ in terms of  $\JCpx$  and $\JCpy$ and combine them according to \eref{pripscheme}. In particular, the case \GO w\AA /\ME{}/NSO above corresponds to $\JCp=\JCpx-\JCpy$. 
 \be{xXY}
\begin{array}{rcl}
%\left( \frac{1}{\ben+\ben\pr} -\frac{1}{\DT \ben} \right)
\JCpx_\uv{\ben} & = & + \int_\vec{k\pr} \frac{\Wkk}{2\pi} \pripm
\sin N\DT \theta f_z   
\\
\JCpy_\uv{\ben} & = & -\int_\vec{k\pr} \frac{\Wkk}{2\pi} \pripm
\sin N\DT \theta f_z\pr ,
\\   
 \JCpx_\uv{\cen} &=&-\int_\vec{k\pr}  \frac{\Wkk}{2\pi}
\left ( 
- \pripm \cos N\Delta \theta f_z+
\pripp f_z\pr 
\right)
 \\ 
 \JCpy_\uv{\cen} &=&-\int_\vec{k\pr}  \frac{\Wkk}{2\pi}
\left ( 
- \pripp  f_z+
\pripm \cos N\Delta \theta f_z\pr 
\right) \\
 \JCpx_z &=& -\int_\vec{k\pr}  \frac{\Wkk}{2\pi}
 \left \{
\pripm
  (
   \sin N \Delta \theta f_\uv{\ben}+\cos N\Delta \theta  f_\uv{\cen} 
 )
 -
 \pripp
 (
 - \sin N  \Delta \theta f_\uv{\ben}\pr +\cos N\Delta \theta  f_\uv{\cen} \pr 
 ) 
 \right \}
 \\
  \JCpy_z &=& -\int_\vec{k\pr} \frac{\Wkk}{2\pi}
\left( \pripp    f_\uv{\cen} - \pripm  f_\uv{\cen} \pr  \right )\, .
\end{array}
\ee

For the same reasons as before only the Fourier components $n=\pm 1$ of the nonequilibrium part $\fee$ can be nonzero. Therefore one only needs to consider
 \be{xXYF}
\begin{array}{rcl}
\JCpx_{\uv{\ben}1} & = & 0   
\\
\JCpy_{\uv{\ben}1} & = & +i\int_\vec{k\pr} \frac{\Wkk}{2\pi}
\pripm
\sin N\DT \theta \sin\DT \theta f_{z1}\pr ,
\\   
 \JCpx_{\uv{\cen}1} &=&-\int_\vec{k\pr}  \frac{\Wkk}{2\pi}
\left( 
- \pripm \cos N \Delta \theta f_{z1}
+
\pripp \cos \Delta \theta f_{z1}\pr
\right)  
 \\
 \JCpy_{\uv{\cen}1} &=&-\int_\vec{k\pr}  \frac{\Wkk}{2\pi}
\left( 
- \pripp  f_{z1}
+
\pripm \cos N \Delta \theta \cos \Delta \theta f_{z1}\pr
\right)  
 \\
 \JCpx_{z1} &=& -\int_\vec{k\pr}  \frac{\Wkk}{2\pi}
\left \{
\pripm 
\cos N\Delta \theta  f_{\uv{\cen}1}
 - 
\pripp
(
i \sin N\Delta \theta \sin \Delta \theta  f_{\uv{\ben}1}\pr + \cos N\Delta \theta \cos  \Delta \theta  f_{\uv{\cen}1} \pr
) 
\right\}
 \\
  \JCpy_{z1} &=& -\int_\vec{k\pr} \frac{\Wkk}{2\pi}
\left(
 \pripp f_{\uv{\cen}1} - \pripm \cos \theta \Delta f_{\uv{\cen}1}\pr 
\right)  
\end{array}
\ee  
Both in \eref{xXY} and in \eref{xXYF} have we left out all terms that vanish due to the assumed symmetry 
$W(-\DT\theta)=W(\DT\theta)$ of the potential.  
We will see below that for the calculation of  the first quantum correction one only  needs to know $\JCp_z[f_\uv{\ben}]$ and $\JCp_\uv{\ben}[f_z]$, that is
\be{jcpxy}
\JCpx_{z1}[f_\uv{\ben}] & =&
 i\int_\vec{k\pr}  \frac{\Wkk}{2\pi}
\pripp 
 \sin N\DT\theta \sin\DT \theta f_{\uv{\ben}1} \pr,
 \nonumber \\
\JCpy_{\uv{\ben}1}[f_z] & = & i\int_\vec{k\pr} \frac{\Wkk}{2\pi}
\pripm 
\sin N\DT \theta \sin\DT \theta f_{z1}\pr  .
\ee

We can  already at this stage extract some general conclusions  for the multilayer case $|N|>1$ with point-like impurities.  The trivial vanishing of the angular integral in  \eref{jcpxy} will imply   below that principal value terms do not contribute to order $(\ell k_\fermi)^0$.   Nor 
do they contribute to order $(\ell k_\fermi)^{-1}$ (see  appendix~\ref{s:secondrecursion}). 
Combining \eref{pripscheme} with the fact   the terms $\JCpy$  in \eref{xXYF}  vanish trivially we conclude that  principal value terms do not contribute to any order for the \GT~approaches.  The integrals $\JCpx$ vanish non-trivially in the special case $|N|=2$, in which case the principal value vanish in all approaches (including the density matrix approach\cite{Trushin-etal-2010}).  To the orders to which principal value terms do not contribute, the quantum corrections are given by the treatment in section~\ref{s:firsttry}. For the monolayer case principal value terms can contribute to the correction $\sim (\ell k_\fermi)^0$  in all approaches, with the trivial exception of \GT w\SKBA.

The Boltzmann equation for $\vec{f}^{(E)}$ (we henceforth drop the the superscript $(E)$) can be written as
\be{sigmaomega}
\vec{\Drive}~=~ \vec{\Precession}[\vec{f}]+\vec{\JCd}[\vec{f}]+\vec{\JCp}[\vec{f}],
\ee  
where $\vec{\Drive}$ is the driving term with 
\be{}
\mtrx{c}{
\Drive_\uv{\ben} \\
\Drive_\uv{\cen} \\
\Drive_z
}=
\mtrx{c}{
E_\uv{k} \partial_k\feq_\uv{\ben} \\
E_\uv{\theta} \tfrac {N}{k}\feq_\uv{\ben} \\
0}\, ,
\ee
$\vec{\Precession}[\vec{f}]$ is the spin-precession term
\be{}
\mtrx{c}{\Precession_\uv{\ben}[\vec{f}] \\ \Precession_\uv{\cen}[\vec{f}] \\  \Precession_z[\vec{f}]
}
=\mtrx{c}{ 0 \\ -2b f_z \\ 2 b f_\uv{\cen}  },
\ee
and the functionals $\JCd$ and $\JCp$ are read off from \eref{bedirackel}, \eref{bediracnso}, \eref{xnso} and the remarks 
below \eref{xnso}. A more informative way of writing the equation \eref{sigmaomega} is 
\be{fullbe}
\begin{array}{rclclcl}
\Drive_\uv{\ben} & = & 0 & + & \JCd_\uv{\ben}[f_\uv{\ben},f_\uv{\cen}] & + & \JCp_\uv{\ben}[f_z] \\
\Drive_\uv{\cen} & = & \Precession_\uv{\cen}[f_z] & + & \JCd_\uv{\cen}[f_\uv{\ben},f_\uv{\cen}] & + & \JCp_\uv{\cen}[f_z] \\
0 &= & \Precession_z[f_\uv{\cen}] & + & \JCd_z[f_z] & + & \JCp_z[f_\uv{\ben},f_\uv{\cen}]. \\
\end{array}
\ee

For notational simplicity we now prefer  to see  the  expansion of $f$ in orders of $(\ell k_\fermi)^{-1}$  as one in powers of $\Wkk$, 
\ie  
\be{}
f~=~f^{(-1)}+f^{(0)}+ f^{(1)}+\ldots 
\ee
with $f^{(n)}\propto W^n\propto (\ell k_\fermi)^{-n}$.
Here $f^{(-1)}\propto W^{-1}$ is the lowest order result that yields the Drude conductivity. Notice that the functionals $\JC[f]$ increase the power in $W$ by one  whereas the action of  $\Precession[f]$ is neutral in powers of $W$. Therefore, the two latter equations do not allow $f_\uv{\cen}$ and $f_z$ to have a lowest order component  $W^{-1}$, since $\Precession[f^{(-1)}]$ would return a term of order $W^{-1}$, which could not be matched by any of the other terms $\Drive$  ($\sim W^0$) and $\JC[f]$ ($\sim W^0$ and higher). The absence
of 
$\Precession_\uv{\ben}$ in the first equation 
(the diagonal components do not precess) is what allows only $f_\uv{\ben}$ to have a term  of order $W^{-1}$. 
Solving $\Drive_\uv{\ben}=\JC_\uv{\ben}[f_\uv{\ben}^{(-1)}]$  ($\sim W^0$), yields $f_\uv{\ben}^{(-1)}=-E_\uv{k}\ttr \partial_k \feq$. 

The components $f^{(0)}$ are found by solving the system
\be{befullW0}
\begin{array}{rclclclr}
0 & = & 0 & + & \JCd_\uv{\ben}[f_\uv{\ben}^{(0)},f_\uv{\cen}^{(0)}] & + & \JCp_\uv{\ben}[f_z^{(0)}] & (\sim W^1) \\
\Drive_\uv{\cen} & = & \Precession_\uv{\cen}[f_z^{(0)}] & + & \JCd_\uv{\cen}[f_\uv{\ben}^{(-1)}] & + & 0  
& (\sim W^0) \\
0 &= & \Precession_z[f_\uv{\cen}^{(0)}] & + & 0& + & \JCp_z[f_\uv{\ben}^{(-1)}] 
& (\sim W^0) \\
\end{array}
\ee
where $f_\uv{\ben}^{(-1)}$ is 
known. The two latter equations constitute a closed set, which  allows 
us to first find  $f_\uv{\cen}^{(0)}$ and $f_z^{(0)}$.  Only the known component $f_\uv{\ben}^{(-1)}$  
goes into the principal value part.  
The system of equation is solved as in the
previous sections by Fourier decomposition and matrix inversion, 
\be{fcfz}
\mtrx{c}{
\frac{iN\mathcal{E}^*}{2k} 
 \feq_\uv{\ben}  
 \\ 0 
 } 
 &~=~&
 \mtrx{cc}{0  & -2b \\ 2b & 0}
 \mtrx{c}{f_{\uv{\cen}1}^{(0)} 
 \\  f_{z1}^{(0)}
 }+
 \mtrx{c}{i\IC^\mathrm{s} f_{\uv{\ben}1}^{(-1)} 
 \\ 0}+
 \mtrx{c}{0 \\ \JCp_{z1}[f_\uv{\ben}^{(-1)} ]} 
  \nonumber\\
 \Rightarrow\quad
  \mtrx{c}{f_{\uv{\cen}1}^{(0)} 
 \\  f_{z1}^{(0)}
 }&~=~& 
 \mtrx{l}{
-\frac{1}{2\ben} \JCp_{z1}[f_\uv{\ben}^{(-1)} ] \\ 
-\frac{1}{2\ben} \frac{i\mathcal{E}^*}{2}  
\left(
\frac{N}{k}\feq_\uv{\ben}+
\frac{
\IC^\mathrm{s}
}{
\IC^+
}
\partial_k\feq_\uv{\ben} 
\right)
 }.
 \ee

Notice that if we discard the principal value terms $\JCp$ we find that $f_\uv{\cen}^{(0)}=0$, which  using 
the first equation in \eref{befullW0} implies $f_\uv{\ben}^{(0)}=0$. This is exactly what the solution \eref{ffromM}  
tells us: there is no $W^{0}$ correction [\ie $(\ell k_\fermi)^0$ correction] to the conductivity in \eref{sigmab}; { only 
 $f_z$ obtains
}  a contribution to order $W^0$. The last line of  \eref{fcfz}  indeed corresponds to  $f_z$ in \eref{ffromM}.

Including principal value terms yields a nonzero $f_\uv{\cen}^{(0)}$. It also gives a nonzero $f_\uv{\ben}^{(0)}$ according to the first equation in \eref{befullW0}, 
\be{}
 0 ~=~ \JCd_\uv{\ben}[f_\uv{\ben}^{(0)},f_\uv{\cen}^{(0)}] + \JCp_{\uv{\ben}1}[f_z^{(0)}]~=~
 -\IC^+ f_{\uv{\ben}1}^{(0)} -i\IC^\mathrm{s}f_{\uv{\cen}1}^{(0)} +  \JCp_{\uv{\ben}1}[f_z^{(0)}]\, . 
\ee
Nonzero in-plane components 
\be{foprip}
f_{\uv{\ben}1}^{(0)} &=&
i\frac{\IC^\mathrm{s}}{2\ben \IC^+} \JCp_{z1}[f_\uv{\ben}^{(-1)}]+ \frac{1}{\IC^+} \JCp_{\uv{\ben}1}[f_z^{(0)}],
\nonumber \\
f_{\uv{\cen}1}^{(0)} &=& -\frac{1}{2\ben} \JCp_{z1}[f_\uv{\ben}^{(-1)}] 
\ee 
result in a correction to the conductivity. In particular, this correction is of order $(\ell k_\fermi)^0$ since the components in \eref{foprip} are of order $(\ell k_\fermi)^0$. 

A closer inspection  shows that 
$f_{\uv{\ben}1}^{(0)}/\mathcal{E}^*$ is real  and $f_{\uv{\cen}1}^{(0)}/ \mathcal{E}^*$ is imaginary, 
as was the case in \eref{ffromM}. This implies [consult equations \eref{feedc} and \eref{E}] that $f_{\uv{\ben}}^{(0)}=E_\uv{k} 2f_{\uv{\ben}1}^{(0)}/ \mathcal{E}^*$ and  $ f_{\uv{\cen}}^{(0)}=E_\uv{\theta} 2 f_{\uv{\cen}1}^{(0)}/ i \mathcal{E}^*$. The first quantum correction $\delta \sigma$  to the conductivity is therefore for arbitrary $N$ given by
\be{}
\vec{E}\delta \sigma~ =~ 2 \int_\vec{k} 
\left(\uv{k}f_{\uv{\ben}}^{(0)}\partial_k \ben  +     \uv{\theta}f_{\uv{\cen}}^{(0)}  \frac{N\ben}{k} \right )
~ =~2\vec{E}\int \frac{k\der k}{2\pi } \left (\frac{f_{\uv{\ben}1}^{(0)} }{ \mathcal{E}^*}\partial_k \ben +
\frac{f_{\uv{\cen}1}^{(0)} }{ i\mathcal{E}^*} \frac{N\ben}{k} 
\right) \propto (\ell k_\fermi)^0 \, .
\ee
From \eref{foprip} one obtains the  first quantum correction as a sum of  the contributions
\be{pripcorrsigma}
\delta \sigma^X &=&- 2  \int \frac{k\der k}{2\pi}  \frac{1}{i\mathcal{E}^* 2\ben }
\left (\frac{\IC^\mathrm{s}}{\IC^+}\partial_k \ben +\frac{N\ben}{k}\right )\JCp_{z1}[f_\uv{\ben}^{(-1)}],
\nonumber  \\
\delta \sigma^Y &=&+2 \int \frac{k\der k}{2\pi} \frac{\partial_k \ben}{\mathcal{E}^* \IC^+ }
\JCp_{\uv{\ben}1}[f_z^{(0)}]\, .
\ee
This can be  written as 
\be{deltasigmageneral}
\delta \sigma^X & =& +
\frac{1}{2\pi}\int \frac{k\der k}{2\pi}
\left(\frac{\IC^\mathrm{s}_k}{\IC^+_k} \partial_k \ben +\frac{N\ben}{k}  \right )
\int \frac{k\pr \der k\pr}{2\pi} 
\frac{1}{\IC^+_{k\pr}}\partial_{k\pr}{\feq_\uv{\ben}}\pr \int_{\theta\pr}  W_{kk\pr\Delta\theta}   \frac{\sin N\Delta \theta \sin \Delta\theta}{\ben^2-{\ben\pr}^2},
\nonumber \\
\delta\sigma^Y & =& -
\frac{1}{2\pi}\int \frac{k\der k}{2\pi}
\frac{1}{\IC^+_k}\partial_k\ben   
\int \frac{k\pr \der k\pr}{2\pi} 
\left(\frac{\IC^\mathrm{s}_{k\pr}}{\IC^+_{k\pr}}  \partial_{k\pr} {\feq_\uv{\ben}}\pr
+\frac{N}{k\pr} {\feq_\uv{\ben}}\pr
\right)
 \int_{\theta\pr} W_{kk\pr\Delta\theta}  \frac{\sin N\Delta \theta \sin \Delta\theta}{\ben^2-{\ben\pr}^2}\, ,
\ee
where we introduced the notation  $W_{kk\pr\Delta\theta}:=\Wkk$.
It is easily seen that for point-like impurities the angular integral vanishes trivially for $|N|\geq 2$.  For further results on point-like impurities see appendix~\ref{s:secondrecursion}. 

At zero temperature the correction \eref{deltasigmageneral} can be written as 
\be{deltasigmageneralT0}
\delta \sigma^X & =& -
\frac{}{}
 \frac{k_\fermi}{8\pi^2 \IC^+_\fermi}
 \int \frac{k\der k}{2\pi}
\left(\frac{\IC^\mathrm{s}_k}{\IC^+_k} \partial_k \ben +\frac{N\ben}{k}  \right )
\int_{\theta\pr} W_{kk_\fermi\Delta\theta} \frac{\sin N\Delta \theta \sin \Delta\theta}{\ben^2-{\ben_\fermi}^2} 
\nonumber \\
\delta\sigma^Y & =& +
\frac{}{}\frac{k_\fermi \IC^\mathrm{s}_\fermi}{8\pi^2 \IC^+_\fermi}
\int \frac{k\der k}{2\pi}
\frac{1}{\IC^+_k}\partial_k\ben   
 \int_{\theta\pr}   W_{kk_\fermi\Delta\theta}  \frac{\sin N\Delta \theta \sin \Delta\theta}{\ben^2-{\ben_\fermi}^2}+
 \nonumber \\
 & & +
\frac{}{}\frac{N}{8\pi^2} \int \frac{k\der k}{2\pi}
\frac{1}{\IC^+_k}\partial_k\ben   
\int_{k_\fermi}^\infty  \der k\pr 
\int_{\theta\pr}  W_{kk\pr\Delta\theta}  \frac{\sin N\Delta \theta \sin \Delta\theta}{\ben^2-{\ben\pr}^2}
\ee
and in the monolayer case  $N=\pm1$ this  can be simplified  
to (with $\partial_k \ben=\tfrac{\ben}{k}=\vf$ and writing $\IC^+_k=\tfrac{k}{4\pi \vf }\int_{ \theta\pr} W_{kk\Delta \theta}\sin^2 \Delta\theta$)
\be{deltasigmaamor}
\delta \sigma^X & =& -
\frac{e^2}{2\pi h}
\frac{ 
\int \der k   \frac{k}{k^2-k_\fermi^2} \int_{\theta\pr}  W_{kk_\fermi \Delta\theta} \sin^2 \Delta \theta   
}{
\int_{\theta\pr }    W_{k_\fermi k_\fermi \Delta\theta} \sin^2 \Delta \theta   
}
\nonumber, \\
\delta\sigma^Y & =& +
\frac{e^2}{2\pi h}
\int \der k   \frac{k_\fermi }{k^2-k_\fermi^2} \frac{ \int_{\theta\pr}  W_{kk_\fermi \Delta\theta} \sin^2 \Delta \theta   
}{
\int_{\theta\pr }    W_{k k \Delta\theta} \sin^2 \Delta \theta   
}+
 \nonumber \\
 & & +
\frac{e^2}{2 \pi h} \int \der k \int_{k_\fermi}^\infty \der k\pr \frac{1}{k^2-{k\pr}^2}  
 \frac{ \int_{\theta\pr}  W_{kk\pr \Delta\theta} \sin^2 \Delta \theta   
}{
\int_{\theta\pr }    W_{k k \Delta\theta} \sin^2 \Delta \theta   
},
\ee
where we reintroduced $e$ and $\hbar$. For point-like impurities these integrals are easily evaluated:
\be{}
\begin{array}{rclcl} 
\delta\sigma^X / \tfrac{e^2}{2\pi h}  &=& - \int _0^{k_\Lambda} \der k \frac{k}{k^2-k_\fermi^2} 
&=&   
- \log \frac{k_\Lambda}{k_\fermi} +\mathcal{O} ((\tfrac{ k_\fermi}{k_\Lambda})^2)
 \, ,

\\
\delta \sigma^Y / \tfrac{e^2}{2\pi h} &=& 
\int _0^{k_\Lambda}\der k \frac{k_\fermi}{k^2-k_\fermi^2} + \int_0^{k_\Lambda} \der k \int_{k_\fermi}^{k_\Lambda}\der k\pr  \frac{1}{k^2-{k\pr}^2} &=&
-\frac{\pi^2}{8} +\mathcal{O} ((\tfrac{ k_\fermi}{k_\Lambda})^2) \, , 
\end{array}   
\ee
where an ultraviolet cutoff $k_\Lambda \gg k_\fermi$ was introduced. Notice that only $\delta \sigma^X$ is ultraviolet divergent. 

The total quantum correction is $\delta \sigma= 4\,(\pm \delta \sigma^X\pm \delta  \sigma^Y)$ with  relative prefactors (possibly zero) given by \eref{pripscheme} for the different approaches.
Notice that the leading quantum correction will not  be positive in all approaches, in contrast to the situation in section~\ref{s:firsttry}. For point-like impurities, in particular, we find 
\be{pripschemepl}
\begin{array}{l l l | l }
&  &  & \delta \sigma /  \tfrac{2e^2}{\pi h } 
 \\
 \hline
&\textrm{\GO}  & \textrm{GKBA} &  - \log \frac{k_\Lambda}{k_\fermi} -\frac{\pi^2}{8}\\

\textrm{\ME\, \& NSO \&}  &\textrm{\GO}  & \textrm{\AA} &  - \log \frac{k_\Lambda}{k_\fermi}+\frac{\pi^2}{8}  \\ \

&\textrm{\GO}  & \textrm{\SKBA} & -  \log \frac{k_\Lambda}{k_\fermi}  \\

&\textrm{\GT}  & \textrm{GKBA} &  -\frac{\pi^2}{8}\\

&\textrm{\GT}  & \textrm{\AA} &  +\frac{\pi^2}{8}\\

& \textrm{\GT}  & \textrm{\SKBA} &  0 

\end{array}
\ee
The leading quantum correction is  ultraviolet divergent  for all the \GO~ approaches, including the density matrix approaches, whereas it is convergent for  the \GT~ approaches. 
Furthermore, only the approach \GT w\AA~ gives a positive correction, namely 
\be{gos}
\delta \sigma = \frac{\pi e^2}{4h} \quad\quad \textrm{(\GT w\AA)}   \, .   
\ee
In the Boltzmann regime $\ell k_\fermi\gg 1 $ this is a small positive shift to the much bigger Drude conductivity $4\sigma_0=\tfrac{2e^2}{h} \ell k_\fermi = \tfrac{e^2}{h} \tfrac{8\vf^2}{\nimp V_0^2}$ for a constant potential $V_{\vec{k}\vec{k\pr}}=V_0$. We mention that for screened charge impurities ultraviolet divergences  are absent.  

To obtain a contribution  of order  $(\ell k_\fermi)^{-1}$,  which should explain the initial onset of convexity in the conductivity as one approaches the Dirac regime, one iterates this recursive procedure, see appendix~\ref{s:secondrecursion}.  
It  is once again the case 
that  one  only needs to insert known distribution functions into the integrals containing principal values. Therefore an analytical solution is possible although increasingly cumbersome. 
By considering the expansion parameter $(\IC^++i\omega)/\ben$ the iterative procedure can be repeated in order 
to find corrections to the ac conductivity. The ac analogue is obtained with the substitutions \eref{acsub}.

The result \eref{gos} applies to point-like impurities. Strictly speaking the assumption of negligible inter-valley scattering should break down, and it is questionable if the results can be used to discuss graphene experiments. This caveat does not apply neither for topological insulators with only  one Dirac cone (see the end of the Introduction), (nor, of course,  
to numerical simulations of graphene including only one cone.   Our main drive, however,  is graphene with screened charged impurities and in particular monolayer graphene with the screening parameter $\qs:=k_\textrm{TF}/k_\fermi \approx 3.2$ relevant for samples on silicon-oxide substrates. We will assume that this is already long-range enough for inter-valley scattering to be of secondary importance. In that case the two-valley results and one-valley results should be roughly the same, and we can discuss the former relying on results for the latter. We can now speculate 
that our leading quantum correction could be one of  the contributions to  
the residual conductivity observed in the experiments of Chen \etal\cite{Chen-Jang-Adam-Fuhrer-Williams-Ishigami-2008}.  We 
plan to return to the quantitative analysis in future work. 
However, already here we can draw some qualitative conclusions based on the present section and  recalling facts from  sec.~\ref{s:model}. 
The value of the correction can only  depend on the dimensionless parameter 
 $\qs$, which for monolayers is  independent of $k_\fermi$ and  hence independent of the electron density. This leads to a rigid vertical shift of  the Drude conductivity as a function of electron density as illustrated in fig.~\ref{f:sketch}.  The size of this shift  depends only  
on natural constants and the dielectric constant present in $k_\mathrm{TF}$. Thus, the quantum correction could depend on the dielectric environment of the monolayer graphene sample.  

We have given a quantitative evaluation for  the correction $(\ell k_\fermi)^0$ in the limit $\qs \rightarrow \infty $ because here  this   limit coincides with that of  point-like impurities. (This was  the situation for  the Drude conductivity (see sec.~\ref{s:model}). However, there  the absolute scaling  with $k_\fermi$ in the  scattering times was relevant, in contrast to the case of the correction \eref{deltasigmaamor}.) We 
have also evaluated the shift  \eqref{deltasigmaamor} in  the opposite limit  $\qs=0<1$ of an unscreened Coulomb interaction. We find the corrections to be ultraviolet divergent and we find the sign of $\sigma^Y$ to be the opposite.    However, since   $\qs \approx 3.2>1 $,  we expect the limit 
$\qs \rightarrow \infty $ to be the more relevant limit. Therefore we expect 
also for the realistic  value $\qs=3.2$ to encounter the case that only the one approach \GT w\AA~gives a positive value and that  value likely to be  close to \eref{gos}. In case inter-valley scattering is negligible with $\qs=3.2$ this value should then also be relevant for the two-valley situation and thus for graphene experiments.

Puddle formation due to charge 
inhomogeneities  
  leads to a variation in the Fermi level, see \eg ref.~\cite{DasSarma-Adam-Hwang-Rossi-review-2010}.  However, our shift should be insensitive at least to small variations as it is independent of the Fermi level.  
In the case where the impurities sit at a non-negligible average distance $d$ from the graphene plane this introduces a second dimensionless parameter $k_\fermi d$, which depends on the density. In this scenario $\delta \sigma$ becomes  density dependent and the shift is no more rigid. However, in the Fermi momentum range $\ell^{-1} \ll k_\fermi \ll d^{-1}$, which is the  range where the Drude conductivity should be linear, the effect on the statements above  should be negligible. Thus, the residual conductivity we make predictions for should be fitted only from the ``strictly" linear part of the Boltzmann conductivity.  

Next we discuss the leading quantum correction \eref{deltasigmageneralT0} in multilayers. We have shown that for point-like impurities it vanishes trivially. Thus---referring to sec.~\ref{s:model}---we expect the correction to vanish close to the Dirac point. However, the interesting limit is the one far away from the Dirac point where \eref{deltasigmageneralT0}  is all what remains of the 
studied quantum corrections. Here, 
it is more relevant to compare with the uncscreened Coulomb interaction. Irrespective of whether or not the correction is finite in this limit, it becomes doubtful  that our analysis is still valid in this regime. The kinetic approach assumes  that the collision time set by the range of the potential is short compared to the  relaxation time  $\ttr$. For point-like impurities this is certainly the case whereas for  an unscreened Coulomb potential it is highly questionable. For a long-range potential it certainly becomes important to take into account  mean-field effects and their renormalizations of the free drift, which we have neglected. 
These issues are discussed in the context of strongly interacting spinless Fermi systems, see ref.~\cite{Lipavsky-Morawetz-Spicka-book}. At the moment we do not know how to generalize these issues to systems with non-trivial spin.  Therefore, we cannot say much about whether there could be non-vanishing effects due to electron-hole coherence for multilayers with charged impurities. 
Since finite effects of electoron-hole coherence far away from the Dirac regime are very counter-intuitive we find it likely that in a proper treatment of the multilayer problem they would vanish at very high densitites.  However, in monolayers we expect such effects to survive,  as we come to next.

 Monolayer graphene  stands out  in many respects and comes with many surprises  compared to multilayer graphene because of the linear dispersion and the unit winding number $|N|=1$. (See also refs.~\cite{Kailasvuori-2009a, Shytov-Mishchenko-Engel-Halperin-2006} on why $|N|=1$ is special.) We saw already in sec.~\ref{s:model} that monoalyers are different to multilayers and 2DEGs when it comes to how the screening depends on the electron density. Therefore we should not straight away discard as unphyscial the finding of  finite effects of electron-hole coherences far away from the Dirac regime, although we expect no such effects in general and in particular not  in multilayers
 In both monolayers and multilayers 
the Fermi surface---and therefore the number of electrons contributing to a nonequilibrium response---grows linearly with $k_\fermi$. A Kubo formula for the conductivity (see \eg eq. (2) in ref.~\cite{Trushin-etal-2010}) disfavors matrix elements between states with a big  energy separation. 
The Zitterbewegung contribution from each electron would therefore   be 
suppressed by the large  energy denominators $1/[\erg^+(k_\fermi) -\erg^-(k_\fermi)]\sim k_\fermi^{-N} $. 
In the case of multilayers this suppression wins over the increasing density of states as $k_\fermi$ increases. However, in the monolayer case the two effects could compensate each other, wherefore a finite effect of Zitterbewegung at large energy splitting is not inconceivable.

 One might worry about the electron-hole coherent effects  being  negligible compared to weak localization corrections. However, this is not necessarily the case, at least in idealized situations, as should be clear from  recent numerics \cite{Trushin-etal-2010}, where the analytically found electron-hole coherent conductivity stays very close to the numerically exact value, with the small rather constant  discrepancy probably due to weak localization. Nor should  the electron-hole coherent shift in monolayers be negligible in the residual conductivity since we find it to be of the order of  one quantum of conductance.  Further,  the different  leading order quantum corrections can be cleanly separated and therefore  treated independently. From the kinetic equation treatment of weak localization in ref.~\cite{Rammer-Smith-1986} we see that the weak localization correction takes only  the Drude response part of the non-equilibrium Green's function as its input and not the full Green's function including  contributions of higher order in $(\ell k_\fermi)^{-1}$. Thus,  like  the Drude response, the weak localization correction should  be independent of the choice of formalism. Therefore, we believe that the weak localization correction and our $(\ell k_\fermi)^0$ correction can be cleanly separated, and that the issue of formalism affects only the latter.

\section{Conclusions and outlook}\label{s:conclusion}
\noindent
In this paper we 
investigated different derivations of semiclassical but spin-coherent Boltzmann equations in a case where differences could matter, namely in the electron-hole coherence originated  quantum corrections to the Drude   conductivity for 2d Dirac electrons, as encountered in 
in graphene or in  the surface states of 3d topological insulators like 
 Bi$_{1-x}$Sb$_x$,   Bi$_2$Te$_3$,  Sb$_2$Te$_3$ and  Bi$_2$Se$_3$.
 With a few exceptions  we find these quantum corrections to be  highly sensitive to the approach. We find the leading quantum correction in monolayer graphene to be particularly interesting as a litmus test, and suggest  that a precise determination of this contribution from numerics or experiments  might single out a unique  approach.  This 
 sensitivity has motivated us to search for an Ansatz that provides the link between a quantum Liouville equation derivation and a Green's function derivation, that we find to differ with existing approaches. The simple structures of the 
derived collision integral in their most general form makes this search unambiguous. We have found the missing link, at least for the case of  impurity interactions in the  lowest Born approximation, and propose a novel Ansatz (\AA) of a simple but counterintuitive form that to our knowledge has not been studied before.  

On a more technical level we pointed out that  the fact that the pseudospin-orbit coupling is the  dominant 
term in the hamiltonian is essential for  the 
differences to become important.  The fact that spin-orbit coupling constitutes the entire kinetic part in the graphene case simplifies the collision integral considerably and makes an analytic solution possible. The analytic treatment  becomes non-trivial  due to  
the presence of principal value terms. We discussed  the physical origin 
of these terms and  explained why one also in our kinetic description should take them seriously. In addition,  we showed how to deal with them for arbitrary scalar impurity potentials, at least for not too long-ranged potentials.  We kept the  winding number of the spin-orbit coupling general  in order to address single layer graphene as well as multilayer graphene. 

We found that the first quantum correction depends both on the chosen formalism as well as on whether or not principal value terms are included.   With principal value terms neglected the leading quantum correction is found to be of order  $(\ell k_\fermi)^{-1}$. When they are included and do not vanish the leading quantum correction is of order $(\ell k_\fermi)^0$.  An electron-hole coherence originated quantum correction  $\sim (\ell k_\fermi)^0$ would be a counterintuitive result as it implies that  electron-hole coherent effects could remain finite even far away from the Dirac regime. We discussed why such a result  in the case of monolayers is not absurd, although surprising. 
In multilayers on the other hand we do not expect such a result, and indeed for point-like impurities the  correction $\sim (\ell k_\fermi)^0$ vanishes trivially. For screened charged impurities one encounters the situation that the potential approaches the opposite limit---the  unscreened Coulomb potential---when one increases the density. Since the kinetic analysis in this paper disregarded renormalization effects of the free drift, our analysis should 
break down when the quasiparticle spends a sizable fraction of its time within the interaction range of the impurities. Therefore, we do not attempt to  evaluate the  correction $\sim (\ell k_\fermi)^0$  for multilayers with  charged impurities in the high density limit where the screening is weak.

We argued  that the shift in monolayers due to electron-hole coherences  
should only depend on the dielectric constant through the dimensionless parameter $\qs$. 
We also argued that the evaluated leading correction  for monolayers with point-like impurities should be closely related to the one relevant  to experiments  since for monolayers on silicon-oxide substrates the screening parameter $\qs\approx 3.2$ is bigger than one and therefore  more related to the limit  $\qs=\infty $  than to the limit   $\qs=0$. 
Such a shift  could be one of the  contributions in the residual conductivity observed in recent experiments   \cite{Chen-Jang-Adam-Fuhrer-Williams-Ishigami-2008}.   
  Our contribution $\delta\sigma$ given in \eref{deltasigmaamor}  
 depends crucially on the approach to deriving collision integrals. With a precise measurement of the residual conductivity and a precise knowledge of other contributions (\eg weak (anti-)localization) that one would need to take into account, monolayer graphene would offer an unprecedented setting for experimentally singling out one  among all the approaches restudied or  introduced in the present paper. In  a  comparison with numerics one would  of course have an even more controlled setting.   We plan to address the actual quantitative analysis and comparison with experiments and numerics   in future work.

 The observed residual conductivity is positive.  If also the  contribution  from electron-hole coherences would  be determined to be positive, this should pick the approach called \GT w\AA~ as the unique alternative. Interestingly, this contains  the Ansatz \AA~that is one of our original contributions. In such a case, the electron-hole coherent effects of graphene would require a kinetic equation that not even in principle could have been derived with existing theory.

The more technical work on  the   Ansatz and  the introduction of the Generalized Kadanoff-Baym Ansatz 
 (GKBA)\cite{Lipavsky-Spicka-Velicky-1986}   was prompted by the study of high-field transport for  spinless electrons, see ref.~\cite{Haug-Jauho-book}.  It would be interesting to study the consequences of our Ansatz on transport beyond linear response. The two Ansatzes certainly differ when electron-hole coherent effects are important, but it is not known to us if there is a difference for spinless electrons in strong electric fields.

As already mentioned in the introduction we believe that all the presented approaches might still have to be refined, in particular for the case of weakly screened impurities, which is relevant for   multilayers with charged impurities an at high fermi momentum. The elaborate literature on Boltzmann transport in spinless electrons offers two clear directions for improvement that we believe could also be important for the quantum corrections due to the pseudospin-orbit interaction. 
\begin{enumerate}
\item A proper  accounting for of all terms that could contribute to first order in the electric field.  For example taking into account that also the noninteracting response functions $\GoR$ and $\GoA$ are modified by the electric field and that the gradient expansion of the self-energy terms to first order includes electric field contributions in a gauge invariant formulation. In the context of spin-orbit interactions this was discussed recently in ref.~\cite{Kailasvuori-2009a}. For electron systems with a trivial spin index these issues have been discussed for more than two decades in the context of high-electric-field transport, see ref.~\cite{Haug-Jauho-book} for a review.  See also refs.~\cite{Zubarev-etal-book, Mahan-book}. 
\item A proper extraction of the quasiparticle part $f$ in the kinetic equation for $\rho$ (or $G^<$) and a proper incorporation of renormalizations of the free drift. In the process one unveils the qualitative difference between the electron distribution function $\rho$ and the quasiparticle distribution $f$. The difference appears as a wave-function renormalization factor and as an extra  term containing principal values. The latter term is related to the quickly decohering off-shell motion from the quasiparticle redressing within the interaction radius.  For single-band electrons these issues have also been discussed for some two decades, for example in the context of  Boltzmann treatments of Fermi systems with strong two-body interactions  (for an extensive  review we refer to ref.~\cite{Lipavsky-Morawetz-Spicka-book}, for the context of impurities, see ref.~\cite{Spicka-Lipavsky-Morawetz-1997}). Here it becomes important to recognize that $\feq=\FDf\neq \rho^\textrm{eq}$.  Only by properly separating out the coherences related to the  quasiparticle redressing can one in a controlled way
extract a Boltzmann equation 
that one solves 
by linearizing around an equilibrium 
described by the Fermi-Dirac distribution $\FDf$.  
\end{enumerate}
It is an open question  as to how one can 
generalize these issues to situations with interband coherences.  However, already at the level of treatment given in the present paper we believe the discussed approach to be very promising for understanding electron-hole coherent effects in the conductivity in graphene, with the recent results in ref.~\cite{Trushin-etal-2010} being one example of this promise.

{\it Acknowledgements}
For discussion and comments we gratefully acknowledge  
M.~I. Katsnelson, K. Morawetz, M.~Auslender, P. Brouwer,  D. Culcer, P. Lipavsk{\'y}, A.-P. Jauho,  T. Nunner and  C. Timm. For the second version of the preprint  we gratefully acknowledge 
 M. Trushin for interactions concerning  the  Boltzmann conductivity for point-like impurities in bilayers.  During the beginning of this work J.K. was supported by the Swedish Research Council. M.L. acknowledges support by the DFG through SPP 1285.

\appendix

\section{Translating evolution operators into Green's functions}\label{s:evogreen}
With  $H_\vec{k}=\aen_\vec{k}+\vec{\sigma}\cdot\vec{\ben}_\vec{k}$ we notice that  
\be{}
 & & \int_0^\infty \der t \, e^{-\eta t} e^{-iH_\vec{k}t} f e^{iH_\vec{k\pr}t} ~ = ~
\int_0^\infty \der t\, e^{-\eta t} e^{-i(\aen-\aen\pr)t} 
\sum_{s} \frac{e^{-is\ben t}+s\vec{\sigma}\cdot \uv{\ben} e^{-is\ben t}  }{2} 
f 
\sum_{s\pr} \frac{e^{is\pr\ben\pr t}+s\pr\vec{\sigma}\cdot \uv{\ben\pr} e^{is\pr\ben\pr t}  }{2} ~=~ 
\nonumber \\
& =& \int_0^\infty \der t\, e^{-\eta t} \sum_{ss\pr} e^{-i(\aen+s\ben-\aen\pr-s\pr \ben\pr )t}  
\sproj f \sproj \pr~ =~
\sum_{ss\pr}\sproj f \sproj \pr
 \frac{1}{\eta +i(\erg-\erg\pr)}~ =~\nonumber \\ 
&= & \int \frac{\der\omega}{2\pi}   \sum_{ss\pr}
    \frac{\sproj }{\omega+i\eta -\erg}f  \frac{\sproj \pr}{\omega-i\eta-\erg\pr } =   \int \frac{\der\omega}{2\pi}   \GoR_\vec{k} f \GoA_\vec{k\pr} \, .
\ee

\section{Details on the anti-ordered Kadanoff-Baym Ansatz}
\label{s:aa}
\noindent
In ref.~\cite{Lipavsky-Spicka-Velicky-1986} the correlator $\GL$ is divided into the auxiliary correlators 
\be{}
\GLr(t_1,t_2) & = & \theta (t_1-t_2) \GL(t_1,t_2)\, , \nonumber \\
 \GLa (t_1,t_2) & =& \theta (t_2-t_1)\GL(t_1,t_2) \, . 
\ee
By acting on $\GLr$ with $(\GR)^{-1}$ from the left, using the generalized Kadanoff-Baym equation \eref{gkbe} and then acting on the result with $\GR$ from the left one arrives at
\be{}
\GLr(t_1,t_2)&=& i\GR(t_1,t_2)\GL(t_2,t_2)+
\int_{t_2}^{t_1} \der t \int_{-\infty}^{t_2} \der t\pr \GR_{t_1t_2}\SR_{tt\pr}\GL_{t\pr t_2}     +\GR_{t_1t_2}\SL_{tt\pr}\GA_{t\pr t_2} \, ,
\nonumber \\
\GLa(t_1,t_2)&=&-  i\GL(t_1,t_1)\GA(t_1,t_2)+
\int_{-\infty}^{t_1} \der t \int_{t_1}^{t_2} \der t\pr \GR_{t_1t_2}\SL_{tt\pr}\GA_{t\pr t_2}     +\GL_{t_1t_2}\SA_{tt\pr}\GA_{t\pr t_2} \, .
\ee 
 We complemented this with the corresponding result for $\GLa$ with $(\GA)^{-1}$ and $\GA$ instead acting from the right. The first terms to the right sum up to the GKBA. The integrals are correction terms that fulfill several natural criteria. 1) On the time-diagonal $t_1=t_2$ they vanish, making the GKBA exact. 2) No integrals stretch to $t=+\infty$, \ie the result respects the causality of the 
Kadanoff-Baym equations. 3) One can derive the same equations for $G^>$, \ie particle-hole symmetry remains. 4) The spectral identity
$\GL+G^>=i(\GR-\GA)$ is still satisfied.

The solution can be used to determine $\GL$ iteratively to the desired precision. Thus, the GKBA can be seen as the first term in an expansion in the interaction strength. However, in ref.~\cite{Lipavsky-Spicka-Velicky-1986} it is noted that 
on top of that the arguments of the self-energies $\Sigma_{tt\pr}$ run over disjoint intervals, which makes the integrals even smaller and relates it to the collision time $\tau_0 (\ll \ttr)$,  the small time  the particles spends within the interaction radius.   

We now  copy this treatment but act with the response functions from the opposite sides  (\ie acting on $\GLr$ with  $(\GR)^{-1}$ from the right using the generalized Kadanoff-Baym equation \eref{gkbe} and then acting on the result with $\GR$ again from the right).  This gives us instead 
\be{}
\GLr(t_1,t_2)&=& i\GL(t_1,t_1)\GR(t_1,t_2)+
\int^{t_1} \der t\pr  
(\GR\SL+\GL\SA-\GLr \SR)_{t_1t\pr} \GR_{t\pr t_2}\,,
\nonumber \\
\GLa(t_1,t_2)&=&-  i\GA(t_1,t_2)\GL(t_2,t_2)+
 \int^{t_2} \der t
 \GA_{t_1t}(\SR\GL+\SL\GA-\SA \GLa)_{t t_2} \, .
 \ee 
The criteria 1)-4) are still satisfied.  Note in particular that the causality is respected. However, the result is a bit more complicated and  this time the   variables $t$ and  $t\pr$ in the self-energies no longer run over  disjoint time intervals. Our conclusion is that the expansion can still be seen as one in  the interaction strength but no longer as one in  the collision time. However,  when comparing Boltzmann equations derived using Green's function techniques with one derived with a Liouville equation approach, it is sufficient to be consistent to the given order of the interaction.

\section{Comparison with the Boltzmann equation of Auslender and Katsnelson}\label{s:auskat}
\noindent
The Boltzmann equation for the Dirac cone K was derived by Auslender and Katsnelson \cite{Auslender-Katsnelson-2007} with the NSO formalism. For comparison we are going to translate our Boltzmann equation into theirs.  (Similar to us they do not consider the terms with  $\delta (\ben+\ben\pr)$.) In ref.~\onlinecite{Auslender-Katsnelson-2007} 
the Boltzmann equation is written for the quantities 
$D:=2f_0-1$, $N:=2f_\uv{\ben}+1$ (not be confused with our winding number $N$ which we have set to 1 in what follows) and $g:=f_z-if_\uv{\cen}$. In terms of these   
we obtain
\be{bediracauskat}
\vec{E}\cdot \partial_\vec{k} D
&=& -\int_\vec{k\pr} \Wkk 
\cos^2 \frac{\DT \theta}{2} \DT D ,
\nonumber \\
 \vec{E}\cdot \partial_\vec{k} N+2\frac{E_\uv{\theta}}{k}  \Im g
&=& -\int_\vec{\pr} \Wkk 
\left \{
\delta (\DT \ben) \left (
\cos^2 \frac{\DT \theta}{2} \DT N
+\sin \DT \theta  \Im g\pr \right)+
\left(
\frac{\sin \DT \theta }{\pi \DT \ben}-
\frac{\sin \DT \theta  }{\pi (\ben+\ben\pr)}
\right) \Re g\pr
\right \},
\nonumber \\
\vec{E}\cdot \partial_\vec{k} g-i\frac{E_\uv{\theta}}{2k}(N-1) -i2\ben g  
&=& -\int_\vec{\pr} \Wkk 
\left \{ 
\delta (\DT \ben) \left (
\cos^2 \frac{\DT \theta}{2} \DT g
-i\frac{\sin \DT \theta}{4}  N\pr 
\right)+  
\right.
\nonumber \\
 & & \left.
 +\left(
\frac{
\frac{\sin \DT \theta}{4} N\pr-
i\cos^2 \frac{\DT \theta}{2}  (g+ g\pr)  
 }{\pi \DT \ben}+
\frac{
\frac{\sin \DT \theta}{4} N\pr-
i\cos^2 \frac{\DT \theta}{2}  (g + {g\pr}^*) }{\pi (\ben+\ben\pr)}
\right) 
\right \} \,. 
 \ee
 Linearizing (\eg $D=D_0+\delta D (E)$) the equations in the electric field and 
comparing with (47), (51) and (52) in ref.~\onlinecite{Auslender-Katsnelson-2007} we recognize all the terms. Apart from  some minor differences in prefactors the main difference is  the $\delta D$ in equation (52) is in our case $\delta N$. In our Boltzmann equation the electron density $f_0$  (\ie~$D$) decouples from the spin density $\vec{f}$ (\ie~$N$ and $g$). This decoupling is also found in ref.~\cite{Culcer-Winkler-2007b}.  

\section{On the ac response with principal value terms neglected}\label{s:acdetails}
 \noindent
 In this appendix we concentrate on the monolayer case $N=\pm1$. 
 With \eref{acsub} the determinant becomes 
\be{}
|M|=4\ben^2 (\IC^+ +i\omega) +(\IC^+ + i\omega  )^2(\IC^\kappa+i\omega )-(\IC^\mathrm{s})^2 (\IC^\kappa+i\omega )=
4\ben^2 (\IC^+ +i\omega ) \left(1+\frac{i\omega (2 \IC^++i\omega)(\IC^\kappa+i\omega)}{4\ben^2 (\IC^+ +i\omega)}\right ) .
\ee
The conductivity derived from \eref{feeac} is  given by
\be{}
\sigma^I (\omega)& = & -\vf \int \frac{k\der k}{2\pi} \frac{1}{|M|} \partial_k \feq_\uv{\ben}  
\left( 4\ben^2 +(\IC^\kappa +i\omega)(2\IC^++i\omega)
\right)\, ,  
\nonumber \\
\sigma^{II} (\omega)& = & -\vf \int  \frac{k\der k}{2\pi} \frac{1}{k|M|} \feq_\uv{\ben}(\IC^\kappa+i\omega)(2\IC^++i\omega)
 \, .
\ee
Under the assumption $\omega, \IC\ll \ben_\fermi\equiv |\mu| $ (here $T=0$) the Fermi surface contribution $\sigma^{I}$ can be expanded in $\omega/|\mu|$ and $\IC/|\mu|$  with   $|M|^{-1}\approx \frac{1}{4\ben^2 (\IC^++i\omega)}(1 -\frac{i\omega(2\IC^++i\omega)(\IC^\kappa+i\omega)}{4\ben^2 (\IC^++i\omega)})$ to yield  the leading order correction
\be{sigmaacI}
\sigma^{I}(\omega)~ \approx~ \sigma_0(\omega) \left ( 1+\frac{\IC^+_\fermi(2\IC^+_\fermi+i\omega)(\IC^\kappa_\fermi+i\omega)}{4|\mu|^2 (\IC^+_\fermi+i\omega)}
 \right)
  \, .
\ee
As already visible in \eref{feeac},  in the pure limit $\IC\rightarrow 0$, there are  no corrections  in $\sigma^{I}$   to $\sigma_0=-i\frac{|\mu|}{4\pi \omega}$.  For $\IC\neq 0$,   $\sigma^{I}$ has, in contrast to the dc results of the last section, a correction of order $(\ell k_\fermi )^0$. However, this correction  only affects 
the inductive part $\Im \sigma^{I}$.  For  $\omega \ll \IC^{+}$ and for $\kappa=+$ we find that $\sigma^{I}(\omega)\approx \sigma_0+\frac{1}{8\pi \ell k_\fermi}-i\frac{\omega}{16\pi |\mu|} $.  
 For the contribution $\sigma^{II}$ the above expansion of $|M|^{-1}$ is allowed since $\ben$ only sweeps the interval $[|\mu|,\infty )$. However, the integral is cumbersome. We  only give  
the result in the clean limit $\IC=0$, 
\be{resigmaII}
\sigma^{II}(\omega) = -i\frac{\omega}{4\pi} \int_0^{\infty} \der \ben \frac{\FD {\ben-\mu  } -\FD {-\ben-\mu}}{4\ben^2-\omega^2} 
&~\stackrel{( T\rightarrow 0)}{=}~&i\frac{1}{32\pi}\ln \frac{2|\mu|+\omega}{2|\mu|-\omega}~\approx ~i \frac{ \omega}{32\pi |\mu|}
\, .
\ee   
 In contrast to the corresponding result for  $\IC\neq 0$, this result does not diverge in the limit $\omega \rightarrow 0$.

\section{Details on a second iteration when principal value terms are included}
\label{s:secondrecursion}
\noindent
In the second  iteration 
the Boltzmann equation \eref{fullbe} reads
\be{befullW1}
\begin{array}{rclclclr}
0 & = & 0 & + & \JCd_\uv{\ben}[f_\uv{\ben}^{(1)},f_\uv{\cen}^{(1)}] & + & \JCp_\uv{\ben}[f_z^{(1)}] & (\sim W^2) \\
0 & = & \Precession_\uv{\cen}[f_z^{(1)}] & + & \JCd_\uv{\cen}[f_\uv{\ben}^{(0)},f_\uv{\cen}^{(0)}] & + & \JCp_\uv{\cen}[f_z^{(0)}]  
& (\sim W^1) \\
0 &= & \Precession_z[f_\uv{\cen}^{(1)}] & + & \JCd_z[f_z^{(0)}] 
& + & \JCp_z[f_\uv{\ben}^{(0)},f_\uv{\cen}^{(0)}] 
& (\sim W^1) \\ \, .
\end{array}
\ee
The driving terms  that were still present in the first  iteration 
---see the equation \eref{befullW0}---are completely absent in the second and higher iterations. 
The components $f_\uv{\ben}^{(0)}$,  $f_\uv{\cen}^{(0)}$ and $f_z^{(0)}$---which determined the correction of order $(\ell k_\fermi)^0$ in the conductivity---are  the known input in the Boltzmann equation \eref{befullW1} from which one then extracts the components  $f_\uv{\ben}^{(1)}$,  $f_\uv{\cen}^{(1)}$ and $f_z^{(1)}$. The latter components determine the correction of  order $(\ell k_\fermi)^{-1}$ in the conductivity. 

We see that generally also the contribution of order $(\ell k_\fermi)^{-1}$ depends on the principal value terms. However,  their contribution vanishes again trivially in the case of  point-like impurities together with  $|N|> 1$,  as  it already happened in the first recursion  with the consequence  that $f_\uv{\ben}^{(0)}$,  $f_\uv{\cen}^{(0)}$ and the contribution $\sim (\ell k_\fermi)^0$ to the conductivity are all 
zero.  For this case  we immediately find  $f_\uv{\cen}^{(1)}=-\Precession_{z1}^{-1}[ \JCd_z[f_z^{(0)}]] =
\frac{\IC^\kappa}{2\ben}f_z^{(0)}$  [see \eref{matrixbe}] independent of principal value terms and therefore consistent with \eref{ffromM}.  
In contrast,  $f_z^{(1)}=-\Precession_{\uv{\cen}1}^{-1}[ \JCp_\uv{\cen}[f_z^{(0)}]]$   will depend on principal value terms.  The pertinent expression 
for \GO wGKBA is given by   
\be{jcpcz}
\JCp_{\uv{\cen}}[f_z] & =&\frac{1}{2\pi}
+\int_\vec{k\pr} \Wkk \left(\frac{\sin^2 \tfrac{N \Delta \theta}{2}} {\DT \ben}+\frac{\cos^2\tfrac{N\Delta\theta}{2} }{\ben+\ben\pr}\right) \Delta f_z
 \, ,
\ee
derived with $\Delta \ben \leftrightarrow (\ben+\ben\pr)$ from the result \eref{xnso} for \GO w\AA/ \ME/NSO. For the  comparison with 
other approaches according to \eref{pripscheme} we decompose \eref{jcpcz} into 
\be{jcpczxy}
{\JCpx_{\uv{\cen}1}}[f_z] & =&\frac{1}{4\pi}\int_\vec{k\pr} \Wkk \left(\frac{1} {\DT \ben}+\frac{1}{\ben+\ben\pr}\right) (f_{z1}-e^{-i\Delta\theta}f_{z1}\pr) 
\nonumber, \\ 
{\JCpy_{\uv{\cen}1}}[f_z] & =&\frac{1}{4\pi}
\int_\vec{k\pr} \Wkk \left(\frac{1 }{\ben+\ben\pr}-\frac{1} {\DT \ben}\right) \cos N\Delta\theta   (f_{z1}-e^{-i\Delta\theta}f_{z1}\pr)\, .  
\ee
The contribution $\JCpx$ does not vanish for $|N|> 1$ whereas $\JCpy$ does. From \eref{pripscheme} we deduce that  principal value terms in the approaches \GO~(but not \GT) can contribute to $f_z^{(1)}$. However, this does not help as this contribution  is removed 
in the last step, 
\be{}
f_\uv{\ben}^{(1)}= -(\JCd_\uv{\ben})^{-1}\left[ \JCd_\uv{\ben}[f_\uv{\cen}^{(1)}] +  \JCp_\uv{\ben}[f_z^{(1)}] \right],
\ee
since $\JCp_\uv{\ben}[f_z^{(1)}]$ for $|N|> 1$ vanishes trivially for the  only nonzero Fourier components $n=\pm1$. 

We have understood that principal value contributions to $f_z$ do not matter.  Only if they appear in $f_\uv{\cen}$ there can be a contribution to the current. 
Inspection of \eref{xnso} shows that the only principal value terms that does not trivially  remove 
the $n=1$ Fourier component  in the case $|N|> 1$ are the term $\JCp_z[f_\uv{\cen}]$ and the 
 already considered term $\JCp_\uv{\cen}[f_z]$. Thus $\JCp_{z}[f_\uv{\cen}^{(1)}]$ could contribute to $f_\uv{\cen}^{(2)}$. In a third recursion, \ie for quantum corrections of the order $(\ell k_\fermi)^{-2}$ it is therefore possible that principal value terms contribute to the conductivity even in the case of point-like impurities in the multilayer case $|N|> 1$. The possibility of leaving out the principal value terms requires therefore further investigation.

\bibliography{/Users/janik/research/bibliographyfolder/Kailasvuori_Bibtexfile.bib}

\end{document}